\documentclass[usenatbib,usegraphicx]{mn2e}
\usepackage{natbib}
\bibpunct{(}{)}{;}{a}{}{,}
\usepackage{graphicx}
\usepackage{psfig}
\usepackage{times}
\usepackage{amsmath}
\usepackage{amssymb}

\title[Quasi-periodic accretion and gravitational waves from oscillating
``toroidal neutron stars'']
{Quasi-periodic accretion and gravitational waves from oscillating
``toroidal neutron stars'' around a Schwarzschild black hole}

\author[Zanotti, Rezzolla and Font]
	{Olindo Zanotti$^{(1)}$, Luciano
	Rezzolla$^{(1),(2)}$, and Jos\'e A. Font$^{(3)}$	\\
								\\
	$^{(1)}$SISSA, International School for Advanced Studies,
	Via Beirut, 2-4 34014 Trieste, Italy			\\
	$^{(2)}$INFN, Department of Physics, University of
	Trieste, Via A. Valerio, 2 34127 Trieste, Italy 	\\
	$^{(3)}$Departamento de Astronom\'{\i}a y Astrof\'{\i}sica,
	Universidad de Valencia, Dr. Moliner 50, 46100 Burjassot
	(Valencia), Spain 
	} 
\pagerange{\pageref{firstpage}--\pageref{lastpage}}
\pubyear{2002}

\begin{document}

\maketitle
\label{firstpage}

\begin{abstract}
We present general relativistic hydrodynamics simulations of constant
specific angular momentum tori orbiting a Schwarzschild black hole. These
tori are expected to form as a result of stellar gravitational collapse,
binary neutron star merger or disruption,
can reach very high rest-mass densities
and behave effectively as neutron stars but with a toroidal topology
(i.e. ``toroidal neutron stars''). Our attention is here focussed on the
dynamical response of these objects to axisymmetric perturbations. We
show that, upon the introduction of perturbations, these systems either
become unstable to the runaway instability or exhibit a regular
oscillatory behaviour resulting in a quasi-periodic variation of the
accretion rate as well as of the mass quadrupole. The latter, in
particular, is responsible for the emission of intense gravitational
radiation whose signal-to-noise ratio at the detector is comparable or
larger than the typical one expected in stellar-core collapse, making
these new sources of gravitational waves potentially detectable. We
discuss a systematic investigation of the parameter space both in the
linear and nonlinear regimes, providing estimates of how the
gravitational radiation emitted depends on the mass of the torus and on
the strength of the perturbation.
\end{abstract}

\begin{keywords}
accretion discs -- general relativity --  hydrodynamics -- 
oscillations -- gravitational waves
\end{keywords} 

\date{Accepted 0000 00 00.
      Received 0000 00 00.}

\section{Introduction}
\label{intro}

	The theory of non-geodesic, perfect fluid, relativistic tori
orbiting a black hole has a long history dating back to fundamental works
of the late 1960s and 1970s \citep{boyer:65, abramowicz:74, fishbone:76,
abramowicz:78, kozlowski:78}. One of the most important results obtained
in this series of investigations was the discovery that stationary
barytropic configurations exist in which the matter is contained within
``constant-pressure'' equipotential surfaces. Under rather generic
conditions, these surfaces can possess a sharp cusp on the
equatorial plane. The existence of this cusp does not depend on the
choice of the specific angular momentum distribution and introduces
important dynamical differences with respect to the standard model of
thin accretion discs proposed by Shakura \& Sunyaev (1973) and later
extended to General Relativity by Novikov \& Thorne (1973).

	The first important difference is that the cusp at the inner edge
of the torus can behave as an effective $\mathrm{L}_{1}$ Lagrange point
of a binary system (although this is really a circle), providing a simple
way in which accretion can take place even in the absence of a shear
viscosity in the fluid. The second important difference is that for a
torus filling its outermost closed equipotential surface (the toroidal
equivalent of a Roche lobe) the mass loss through the cusp may lead to a
``runaway'' instability \citep{abramowicz:83}. Any amount of material
accreted onto the black hole through the cusp would change the black hole
mass thus affecting the equipotential surfaces and the location of the
cusp. If the cusp moved to {\it smaller} radial positions, the new
configuration would be of equilibrium and no further accretion would
follow. If, on the other hand, the cusp moved to {\it larger} radial
positions, the new configuration would {\it not} be of equilibrium and
additional material (which was previously in equilibrium) would accrete
onto the black hole which, in turn, would further increase its mass. This
process could trigger a runaway mechanism in which more and more mass is
accreted onto the hole, evacuating the whole torus on a dynamical
timescale between 10 ms and 1 s.

	The runaway instability has also attracted attention in
connection with models of $\gamma$-ray bursts~\citep{daigne:97,
meszaros:02}. In these models, in fact, the central engine is assumed to
be a torus of high density matter orbiting a stellar mass black hole,
with intense electromagnetic emission processes lasting up to a few
seconds (see \citet{ruffert:99,ruffert:01} for a recent review on
this). While many of the investigations of the runaway instability have
concentrated on the stability properties of stationary models (both in
Newtonian gravity and in General Relativity), a time-dependent and fully
relativistic study of the runaway stability has been presented only
recently~\citep{font:02}. Through a number of hydrodynamical simulations,
\citet{font:02} were able to show that, for {\it constant} specific
angular momentum tori slightly overflowing their Roche lobes, the runaway
instability does take place and for a wide range of ratios between the
mass of the torus and that of the black hole.

	It should be noted, however, that while the instability seems a
robust feature of the dynamics of constant specific angular momentum
tori, its existence has been severely questioned under more generic
initial conditions. Different works, in fact, have shown that a more
detailed modelling of the initial configurations can either suppress or
favour the instability. Taking into account the self-gravity of the torus
seems to favour it \citep{masuda:97}. The inclusion of rotation of the
black hole, on the other hand, has a general stabilizing effect
\citep{wilson:84,abramowicz:98}. The same applies for tori with
non-constant angular momentum distributions, as shown first by
\citet{daigne:97} using stationary models and by \citet{masuda:98} with
SPH time-dependent simulations with a pseudo-Newtonian potential. We note
that very recently \citet{font2:02} have extended their relativistic
simulations to the case of non self-gravitating tori with {\it
non-constant} angular momentum, finding that the runaway instability can
be suppressed already with a slowly increasing specific angular momentum
distribution. A summary of the different results obtained with the
different approximations made so far can be found in Table~I of
\citet{font:02}.

	A first aim of this paper is to establish how sensitive the onset
of the instability is on the choice of constant specific angular momentum
configurations that are initially overflowing their Roche lobe. To do
this we adopt a mathematical and numerical approach similar to the one
used by \citet{font:02}. It should be noted, however, that while we
analyze the behaviour of tori that have masses comparable to the ones
considered by \citet{font:02}, they are also much more compact. As a
result, our tori generically have higher rest-mass densities, in some
cases almost reaching nuclear matter density. A second and most important
aim of this paper is to investigate the dynamical response of these
relativistic tori to perturbations.  Our interest for this has a simple
justification: because of their toroidal topology, these objects have
intrinsically large mass quadrupoles and if the latter are induced to
change rapidly as a consequence of perturbations, large amounts of
gravitational waves could be produced.

	Both of the aspects mentioned above justify in part our choice of
terminology. Over the years, in fact, different authors have referred to
these models in a number of ways, starting from the original suggestion
of \citet{abramowicz:78} of ``toroidal stars'' to the more recent and
common denomination of ``accretion tori'', or ``thick discs''. Hereafter,
however, we will refer to these specific objects as tori, but also,
reviving the original definition by~\citet{abramowicz:78}, as {\it
``toroidal neutron stars''}. There are three reasons for this
unconventional choice.  Firstly, these objects have equilibrium
configurations with (small) finite sizes that are pressure supported and
not accreting. In this sense, they are very different and
should not be confused with standard
accretion discs that are in principle infinitely
extended, are generically thin because
not pressure supported and are, of course, accreting. Secondly, these objects
have rest-mass densities much larger than the ones usually associated
with standard accretion discs.
Thirdly, while possessing a toroidal topology these objects
effectively behave as the more familiar neutron stars, most notably in
their response to perturbations.

	While this analogy is attractive, important differences exist
between toroidal and ordinary neutron stars, the most important being
that toroidal neutron stars are generically unstable while spherical
neutron stars are generically stable. More of these differences will
appear in the following Sections. As a final remark we note that the idea
of toroidal neutron stars might appear less bizarre when  considering a
neutron star as a fluid object whose equilibrium is mainly determined by
the balance of gravitational forces, pressure gradients and centrifugal
forces. In this framework then, the familiar neutron stars with spherical
topology are those configurations in which the contributions coming from
the centrifugal force are much smaller than the ones due to  pressure
gradients and  gravitational forces. On the other hand, when the
contributions of the pressure gradients are smaller than the
ones due to the centrifugal and gravitational forces, a toroidal topology
is inevitable and a toroidal neutron star then becomes an obvious
generalization (see \citet{ansorg:02} for a recent summary of the
research on uniformly rotating axisymmetric fluid configurations).

	The plan of the paper is as follows. In Section \ref{analytic} we
briefly recall the main properties of toroidal neutron stars, while
Section \ref{numerical} is devoted to a discussion of the equations to be
solved and of the numerical approach used to solve them. After discussing
our choice of initial data in Section \ref{initialdata}, the results of
the numerical calculations will be presented in Section \ref{results},
reserving Section \ref{gwe} to the discussion of the gravitational
radiation produced. Finally, Section \ref{concl} contains our conclusions
and the plans in which our future research will be organized.

	Throughout, we use a space-like signature $(-,+,+,+)$ and a
system of geometrized units in which $G = c = 1$. The unit of length is
chosen to be the gravitational radius of the black hole, $r_{\rm g}
\equiv G M/c^2$, where $M$ is the mass of the black hole. When useful,
however, cgs units have been reported for clarity. Greek indices are
taken to run from 0 to 3 and Latin indices from 1 to 3.

\section{Analytic solutions for stationary configurations}
\label{analytic}

	In what follows we recall the basic properties of stationary
toroidal fluid configurations in a curved spacetime and the interested
reader will find a more detailed discussion in \citet{font:02}. The
considerations made here will be useful only for the construction of the
background initial model which we will then perturb as detailed in
Section \ref{initialdata}.

	Consider a perfect fluid with four-velocity ${\bf u}$ and
described by the stress-energy tensor
\begin{equation}
\label{stress-tensor}
T^{\mu\nu}\equiv (e+p)u^\mu u^\nu+p g^{\mu\nu}
	= \rho h u^\mu u^\nu+p g^{\mu\nu} \ ,
\end{equation}
where $g^{\mu\nu}$ are the coefficients of the metric which we choose to
be those of a Schwarzschild black hole in spherical coordinates
$(t,r,\theta,\phi)$. Here, $e$, $p$, $\rho$, and $h = (e+p)/\rho$ are the
proper energy density, the isotropic pressure, the rest mass density, and
the specific enthalpy, respectively. In the following we will model the
fluid as ideal with a polytropic equation of state (EOS) $p=\kappa
\rho^\gamma=\rho\epsilon(\gamma-1)$, where $\epsilon=e/\rho - 1$ is the
specific internal energy, $\kappa$ is the polytropic constant and
$\gamma$ is the adiabatic index. Also, for simplicity we will consider
the fluid not be magnetized. This may represent a crude approximation
given that toroidal neutron star are probably created by material
originally magnetized and that very large magnetic fields can be easily
produced when rapid shearing motions are present in highly conducting
magnetized fluids \citep{spruit:99,rls:00}.

	The fluid is assumed to be in circular non-geodesic motion with
four-velocity $u^{\alpha} = (u^t,0,0,u^{\phi}) = u^t(1,0,0,\Omega)$,
where $\Omega = \Omega(r,\theta) \equiv u^{\phi}/u^t$ is the coordinate
angular velocity as observed from infinity. If we indicate with $\ell$
the specific angular momentum\footnote{Note that this is not the only
definition for the specific agular momentum used in the
literature. Often, in fact, the specific angular momentum is defined as
$\ell' \equiv u_\phi$ because this is a constant of geodesic
(i.e. zero-pressure) motion in axially symmetric spacetimes. When the
pressure is non-zero, on the other hand, $h u_{\phi}$ is a constant of
motion, while $\ell'$ is not. For axially symmetric, stationary
spacetimes $\ell \equiv - u_\phi/u_t$ is constant for both geodesic and
perfect fluid motion.}  (i.e. the angular momentum per unit energy) $\ell
\equiv -u_{\phi}/u_t$, the orbital velocity can then be written in terms
of the angular momentum and of the metric functions only, as
\hbox{$\Omega = - \ell (g_{tt}/g_{\phi\phi})$}.

	From the normalization condition for the four-velocity vector,
$u^{\alpha}u_{\alpha}=-1$, we derive both the total specific energy of
the fluid element, $u_t$ and the redshift factor, $u^t$ as
\begin{equation}
\label{energy}
(u_t)^2=-\frac{g_{\phi\phi}g_{tt}}{g_{\phi\phi} 
	+ \ell^2 g_{tt}} \ , \hspace{1cm}
(u^t)^2 = -\frac{1}{g_{tt} + \Omega^2 g_{\phi\phi}}\ .
\end{equation}

	Under these assumptions, the equations of motion for the fluid
can be generically written as $h^i_{\ \nu} \nabla_{\mu} T^{\mu \nu}=0$
where $h_{\mu \nu} \equiv g_{\mu \nu} + u_{\mu}u_{\nu}$ is the projector
tensor orthogonal to ${\bf u}$ and $\nabla$ the covariant derivative in
the Schwarzschild spacetime. Enforcing the conditions of hydrostatic
equilibrium and of axisymmetry simplifies the above equations
considerably. Furthermore, if the contributions coming from the
self-gravity of the torus can be neglected, the relativistic hydrodynamics
equations reduce to Bernoulli-type equations
\begin{equation}
\label{bernoulli}
\frac{\nabla_i p}{e+p} = - \nabla_i W +
	\frac{\Omega \nabla_i \ell}{1- \Omega \ell} \ ,
\end{equation} 
where $i=r,\theta$ and $W = W(r,\theta) \equiv \ln(u_t)$.
	
	The simplest solution to equations (\ref{bernoulli}) is the one
with $\ell = {\rm const.}$, since in this case the equipotential surfaces
can be computed directly through the metric coefficients and the value of
the specific angular momentum. Note that at any point in the $(r,\theta)$
plane, the potential $W$ can either be positive (indicating equipotential
surfaces that are open) or negative (indicating equipotential surfaces
that are closed). The case $W=0$ refers to that special equipotential
surface which is closed at infinity. Interestingly, closed equipotential
surfaces contain local extrema and in the equatorial plane these mark two
very important points. There, in fact, $\nabla_i W = 0 =\nabla_i p$ and
an orbiting fluid element would not experience any net acceleration, with
the centrifugal force balancing the gravitational one exactly. These
points correspond to the (radial) positions of the cusp, $r_{\rm{cusp}}$,
and of the ``centre'' of the torus, $r_{\rm c}$. At these radial
positions the specific angular momentum must be that of a Keplerian
geodesic circular orbit
\begin{equation}
\label{keplerian}
\ell^2 = \ell^2_K \equiv \frac{M r^3}{(r-2M)^2} \ ,
\end{equation} 
which can effectively be used to calculate the position of both the centre 
and the cusp.

	In the case of a torus with constant specific angular momentum, it
is straightforward to show that the position of the cusp is located
between the marginally bound circular orbit, $r_{\rm b}$, and the
marginally stable circular orbit, $r_{\rm ms}$, of a point-like particle
orbiting the black hole \citep{abramowicz:78}. Note that the position of
the inner edge of the torus $r_{\rm in}$ and the position of the cusp
$r_{\rm{cusp}}$ need not coincide and indeed $r_{\rm{in}}$ can be chosen
to be anywhere between the cusp and the centre. Once made, however, the 
choice for $r_{\rm{in}}$ also determines the position of the outer edge 
of the torus on the equatorial plane through the constraint that both points 
belong to the same equipotential surface, i.e. 
$W(r_{\rm{out}},\pi/2) = W(r_{\rm{in}},\pi/2)$.

	A particularly attractive feature of tori with constant specific
angular momentum is that once the background spacetime and the value of
the specific angular momentum have been fixed, the angular velocity
$\Omega=\Omega(r,\theta)$ is fully determined. Furthermore, if a
polytropic EOS is used, the Bernoulli equations (\ref{bernoulli}) can be
integrated analytically to yield the rest-mass density (and pressure)
distribution inside the torus as
\begin{equation}
\label{density}
\rho(r,\theta) = \left\{\frac{\gamma-1}{\kappa \gamma}
	\left[\exp({W_\mathrm{in}-W})-1\right]\right\}^{1/(\gamma-1)} \ ,
\end{equation}
where $W_{\rm in} \equiv W(r_{\rm in},\pi/2)$. Once the rest-mass
distribution is known, the total rest-mass of the torus can be easily
calculated as
\begin{equation}
M_{{\rm t},*} \equiv \int\rho\sqrt{-g}u^t d^3x \ , 
\end{equation}
where $\sqrt{-g}=r^2\sin\theta$, while the total mass-energy in the toroidal 
neutron star is computed as
\begin{equation}
M_{\rm t} \equiv \int (T^r_r + T^\phi_\phi +
	T^\theta_\theta - T^t_t)\sqrt{-g}\,d^3x  \ ,
\end{equation}
where $d^3 x=dr \, d\theta\, d\phi$ is the coordinate volume element.
(Note that for simplicity, hereafter, we will refer to the mass-energy of
the toroidal neutron star as the ``mass'' of the toroidal neutron star.)

	Depending on the value of $\ell$ chosen and in particular on how
this compares with the specific angular momenta corresponding to orbits
that are marginally bound, $\ell_{\rm mb} = 4$, or that are marginally
stable, $\ell_{\rm ms} = 3\sqrt{3/2}$, different configurations can be
built. A detailed classification of the models can be found in the
literature \citep{abramowicz:78, font:02}; here we simply recall that if
$\ell_{\rm ms} < \ell < \ell_{\rm mb}$, there will be an equipotential
surface closed at a finite radius and possessing a cusp. As a result, a
stationary toroidal neutron star of finite extent can be built and this
will represent our fiducial unperturbed toroidal neutron star.

\section{Mathematical Framework}
\label{numerical}

\subsection{Hydrodynamic equations}
\label{he}

	To preserve the conservative nature of the equations of general
relativistic hydrodynamics, namely the local conservation of baryon
number and energy-momentum, it is convenient to cast them in the form of
a flux-conservative hyperbolic system through the introduction of
suitable {\it ``conserved''} variables rather than in terms of the
ordinary fluid, or {\it ``primitive''}, variables. In this case, the
equations assume the form \citep{banyuls:97,font:98a}
\begin{eqnarray}
\label{fcf}
\frac{\partial {\bf U}({\bf w})}{\partial t} +
	\frac{\partial [\alpha {\bf F}^{r}({\bf w})]}{\partial r} +
	\frac{\partial [\alpha {\bf F}^{\theta}({\bf w})]}{\partial \theta}
	 = {\bf S}({\bf w}) \ ,
\label{system}
\end{eqnarray}
where $\alpha=\sqrt{-g_{00}}$ is the lapse function of the Schwarzschild
metric and where ${\bf U}({\bf w})=(D, S_r, S_{\theta}, S_{\phi})$ is the
state-vector of the evolved variables. The other vectors ${\bf F}^{i}$
and ${\bf S}$ appearing in (\ref{fcf}) represent the fluxes and sources
of the evolved quantities, respectively. The relation between the
conserved and primitive variables in the vector ${\bf w} = (\rho, v_{i},
\epsilon)$ are given through the following set of equations
\begin{eqnarray}
\label{evolved}
D &\equiv&  \rho \Gamma \ ,
\nonumber \\
S_j &\equiv&  \rho h \Gamma^2 v_j \ ,
\end{eqnarray}
supplemented with the ideal-fluid EOS. Note that the covariant components
of the three-velocity are defined in terms of the spatial 3-metric
$\gamma_{ij}$ to be $v_i=\gamma_{ij}v^j$, where $v^i=u^i/\alpha
u^t$. (Although in axisymmetry, we evolve also the azimuthal component of
the equations of motion, so that the index $j$ in Eqs. (\ref{evolved})
takes the values $j=r,\theta,\phi$.) The Lorentz factor $\Gamma$ measured
by a local static observer and appearing in Eqs. (\ref{evolved}) is
defined as $\Gamma\equiv\alpha u^t = (1-v^2)^{-1/2}$, with $v^2 \equiv
\gamma_{ij}v^i v^j$. The specific expressions for the components of the
flux vectors ${\bf F}^{i}$ and of the source vector ${\bf S}$ can be
found in \citet{font:02}.

\subsection{Spacetime evolution}
\label{ste}

	The general relativistic hydrodynamics equations we solve assume
that the fluid moves in a curved spacetime (provided by the Schwarzschild
solution) that is static. The onset and development of the runaway
instability, on the other hand, depends crucially on the response of the
fluid to variations of the spacetime and in particular of its
longitudinal part. To follow this in a self-consistent manner would
require the solution of the Einstein field equations together with those
of relativistic hydrodynamics. Computationally, however, this is a much
harder problem. Fully numerical relativity codes evolving black hole
spacetimes with perfect fluid matter (either in two or three spatial
dimensions) are being developed only recently~\citep{brandt:00,
shibata:00, font-goodale:02}. In addition to this, the need of high grid
resolution and of computational timescales that are much larger than the
dynamical one, may still be too taxing for full numerical relativity
codes.

	To avoid the solution of the full Einstein equations and yet
simulate the onset and development of the instability we follow
the approach proposed by \citet{font:02}. Most notably, at each timestep
we calculate the accretion rate at the innermost radial point $r_{\rm
min}$ of the grid as
\begin{equation}
\dot{m}(r_{\rm min})\equiv - 2\pi\int_0^{\pi} \sqrt{-g} D v^r d\theta 
	\Big\vert_{r_{\rm min}}\ ,
\end{equation}
and thus determine the amount of matter accreted onto the black hole as
\begin{equation}
M^{n+1}=M^n + \Delta t \ \dot m^n(r_{\rm min}) \ ,
\end{equation}
where the upper indices refer to a given time-level. Once the new mass of
the black hole has been computed, the relevant metric functions are
instantaneously updated as
\begin{equation}
\label{met_up}
g_{\mu \nu}(r,M^{n}) \longrightarrow {\tilde
	g}_{\mu \nu}(r,M^{n+1}) \ ,
\end{equation}
so that ${\tilde g}_{\mu \nu}$ describes the spacetime at the time-level
$n+1$, over which the hydrodynamical equations will be solved. We note
that to be consistent the transfer of angular momentum from the torus to
the black hole should also be taken into account. While we have not
considered this here, the interested reader will find the details on a
procedure to account for the angular momentum transfer onto the black
hole in~\citet{font2:02}.

	Our approach for the spacetime evolution is clearly an
approximation and it masks important features such as the response of the
black hole to the accreted mass and the corresponding emission of
gravitational radiation.  Nevertheless, this approximation is often very
good especially when the toroidal neutron stars are not very massive and
the rest-mass accretion rates are therefore small. In these cases, then,
the fractional variation of the black hole mass between two adjacent
time-levels is minute and the spacetime evolution can be treated
effectively as a discrete sequence of stationary spacetimes.

\subsection{Numerical approach}
\label{na}

	The numerical code used in our computations is based on a code
that as has been first described in~\citet{font:98a} (see also
\citet{font:02}). This code performs the numerical integration of system
(\ref{system}) using upwind high resolution shock-capturing (HRSC)
schemes based on approximate Riemann solvers (see, e.g. \citet{fontlr}
and references therein). Exploiting the flux conservative form of
equations (\ref{fcf}), the time evolution of the discretized data from a
time-level $n$ to the subsequent one $n+1$ is performed according to the
following scheme
\vbox{
\begin{eqnarray}
\label{godunov}	
{\bf U}_{i,j}^{n+1} = {\bf U}_{i,j}^{n}
	&-&  \frac{\Delta t}{\Delta r}
	\left(\widehat{{\bf F}}^r_{i+1/2,j}-
	      \widehat{{\bf F}}^r_{i-1/2,j}\right)
\nonumber \\ \nonumber \\ 
	&-& \frac{\Delta t}{\Delta \theta}
    	\left(\widehat{{\bf F}}^{\theta}_{i,j+1/2}-
  	      \widehat{{\bf F}}^{\theta}_{i,j-1/2}\right)
\nonumber \\ \nonumber \\ 
	&+& \Delta t \,\, {\bf S}_{i,j} \ ,
\end{eqnarray}}
\noindent
where the subscripts $i,j$ refer to spatial ($r,\theta$) grid points, so
that ${\bf U}_{i,j}^n \equiv {\bf U}(r_i, \theta_j, t^n)$.  The inter-cell
numerical fluxes, $\widehat{{\bf F}}^{r}_{i \pm 1/2,j}$ and
$\widehat{{\bf F}}^{\theta}_{i,j \pm 1/2}$, are computed using Marquina's
approximate Riemann solver (Donat \& Marquina, 1996). A piecewise-linear
cell reconstruction procedure provides second-order accuracy in space,
while the same order in time is obtained with a conservative two-step
second-order Runge-Kutta scheme applied to the above time update.

	Our computational grid consists of $N_r \times N_{\theta}$ zones
in the radial and angular directions, respectively, covering a
computational domain extending from $r_{\rm {min}}=2.1$ to $r_{\rm
{max}}=30$ and from $0$ to $\pi$. We have used numerical grids with
different number of zones, finding that the truncation error is reduced
to satisfactory values when $N_r=250$ and $N_{\theta}=84$. All of the
results presented in the paper have been computed with this number of
grid-points. The radial grid is logarithmically spaced in term of a
tortoise coordinate $r_* = r + 2M \ln (r/2M -1 )$, with the maximum
radial resolution at the innermost grid being $\Delta r=6\times 10^{-4}$.
As in~\citet{font:02}, we use a finer angular grid in the regions that
are usually within the torus and a much coarser one outside. A grid-point
belongs to the external surface of the initial unperturbed torus when
$u_t(r,\theta) = u_{t,\rm{in}} \equiv u_t(r_{\rm{in}},\pi/2)$. This
equation defines the meridional section of the surface as a closed polar
curve ${\cal C}$ of equation
\begin{equation}
\sin\theta=\left[\frac{u_{t,{\rm in}}^2 \ell^2
	(1-2M/r)}{r^2(u_{t,{\rm in}}^2 - 1 + 2M/r)}\right]^{1/2}
	\ .
\end{equation}
The angular extension $\theta_m$ of the unperturbed torus can then be
computed by searching for the local extrema of the curve ${\cal C}$. As a
result, in most of our simulations $75\%$ of the angular grid points are
uniformly distributed in the range $[\theta_m, \pi-\theta_m]$, while the
remaining points cover the external region.

	The boundary conditions adopted, the treatment of the vacuum
region outside the torus with a low density atmosphere, and the procedure
for recovering physical variables from the conserved quantities $D$ and
$S_{i}$ are the same as those used by \citet{font:02}. The interested
reader is referred to that work for further details.

\section{Initial data}
\label{initialdata}

	Simulating a dynamical instability with a numerical code brings
up the problem of suitable initial conditions. A natural choice would be
that of a configuration that is in equilibrium, where the latter is just
a marginally stable one. In this case, then, any perturbation would move
the configuration away from the equilibrium, inducing the instability on
a finite timescale. While we see this happen regularly in Nature, it is
rather difficult to simulate it numerically, the major obstacle being the
need of performing the numerical simulations on those (short) timescales
that can be afforded computationally. Fortunately, however, there are
ways of by-passing this limitation and these generally consist of
choosing an initial configuration which is already slightly out of
equilibrium. By controlling the deviation away from the equilibrium in
some parametrized form, the timescale for the development of the
instability can then be reduced to values that are compatible with the
computational ones.

	An approach of this type has been used in the past also to
simulate the runaway instability, where a measure of deviation away from
the unstable equilibrium was made in terms of the potential difference
$\Delta W_{\rm in}>0$ at the inner edge of the torus. This quantity,
defined as $\Delta W_{\rm in} \equiv W_{\rm {in}} - W_{\rm{cusp}}$,
accounts for the potential jump on the equatorial plane between the inner
edge of the torus and the cusp \citep{igumenshchev:97}. By simply varying
the value of $\Delta W_{\rm in}$, it is then possible to select a
configuration corresponding to a torus inside its Roche lobe and for
which no mass transfer is possible (i.e. $\Delta W_{\rm in}<0$), or a
torus overflowing its Roche lobe and therefore accreting onto the black
hole (i.e. $\Delta W_{\rm in}>0$). The case limiting the two classes of
solutions, (i.e. $\Delta W_{\rm in}=0$) refers to a configuration that is
just {\it marginally stable} to the runaway instability, which will
therefore develop over an infinite timescale. (Note that this condition
is also equivalent to setting $r_{\rm in} = r_{\rm cusp}$.)

\begin{table*}
\begin{center}
\caption{Main properties of the constant angular momentum toroidal
neutron star models used in the numerical calculations. From left to
right the columns report: the name of the model, the star-to-hole mass
ratio $M_\mathrm{d}/M$, the polytropic constant $\kappa$, the specific
angular momentum $\ell$ (normalized to $M$), the inner and outer radii of
the toroidal neutron star $r_\mathrm{in}$ and $r_\mathrm{out}$, the
radial position of cusp $r_\mathrm{cusp}$, the radial position of the
centre $r_\mathrm{centre}$ (all radii are in units of the gravitational
radius $r_\mathrm{g}$), and the orbital period at the centre of the torus
$t_\mathrm{orb}$, expressed in milliseconds.  The last two columns
indicate the density at the centre of the torus and the average density
of each model, respectively, both in cgs units. All of the models share
the same mass for the black hole, $M=2.5M_{\odot}$ and adiabatic index
$\gamma=4/3$.}
\label{tab1}
\begin{tabular}{l|l|cc|cc|ccc|ccc}
\hline
Model   & $M_{\rm t}/M$    & $\kappa$      &
$\ell$  & $r_{\rm in}$     & $r_{\rm out}$ & $r_{\rm cusp}$ 
        & $r_{\rm centre}$ & $t_{\rm orb}$ & $\rho_{\rm centre}$
	& $\langle\rho\rangle$  \\
& & ${\rm (cgs)}$ & & & & & & ${\rm (ms)}$ & ${\rm (cgs)}$ &${\rm (cgs)}$ &\\

\hline
(a)   & 1.     & 4.46${\times} 10^{13}$ &  3.8000  &4.576 &
	15.889 & 4.576  & 8.352 & 1.86 &1.14${\times} 10^{14}$ &4.72${\times}
	10^{12}$  \\
(b)   & 0.5    & 5.62${\times} 10^{13}$ &  3.8000  &4.576 &
	15.889 & 4.576  & 8.352 & 1.86 &5.72${\times} 10^{13}$ &2.36${\times}
	10^{12}$   \\
(c)   & 0.1    & 0.96${\times} 10^{14}$ &  3.8000  &4.576 &
	15.889 & 4.576  & 8.352 & 1.86 &1.14${\times} 10^{13}$ &4.73${\times}
	10^{11}$ \\
(d)   & 0.05   & 1.21${\times} 10^{14}$ &  3.8000  &4.576 &
	15.889 & 4.576  & 8.352 & 1.86 &5.73${\times} 10^{12}$&2.36${\times}
	10^{11}$ \\
\hline
(e)   & 0.1    &   7.0${\times} 10^{13}$ & 3.7845 &4.646 &
	14.367 &4.646 & 8.165 & 1.80 &1.61${\times} 10^{13}$ &6.43${\times}
	10^{11}$  \\
(f)   & 0.1    &   1.0${\times} 10^{14}$ & 3.8022 &4.566 &
	16.122 &4.566 & 8.378 & 1.87 &1.10${\times} 10^{13}$ &4.48${\times}
	10^{11}$ \\
(g)   & 0.1    &   2.0${\times} 10^{14}$ & 3.8425& 4.410
	&21.472 &4.410 &8.839 &2.03  &4.96${\times} 10^{12}$ &2.01${\times}
	10^{11}$  \\
(h)   & 0.1    &   3.5${\times} 10^{14}$ & 3.8800& 4.290
	&29.539 &4.290 &9.246 &2.17  &2.41${\times} 10^{12}$ &8.12${\times}
	10^{10}$\\
\hline
\end{tabular}
\end{center}
\end{table*}

	All of the models considered by \citet{font:02} have been
constructed with potential differences $\Delta W_{\rm in}>0$, so that the
outermost potential surface is not closed at the cusp but reaches the
black hole. After truncating the torus at $r = r_{\rm cusp}$, the
simulations were carried out by evolving the set of equations discussed
in Section \ref{he}. With this choice, a small fraction of the initial
fluid configuration (i.e. all the one residing outside the Roche lobe) is
out of equilibrium. Of course, this is not the only way of triggering the
instability. In the calculations by \citet{masuda:97} the size of the
torus was expanded by a small amount to overflow its Roche lobe and to set
the configuration out of equilibrium. While it has been argued that the
occurrence of the instability is not much affected by the choice of the
initial model \citep{masuda:97} we have here followed a different
approach to the problem of initial condition for the runaway instability
as discussed below.

\subsection{Introducing a perturbation}
\label{iap}

	An important difference with respect to earlier works in our
prescription of the initial data is that we have considered models with a
potential barrier $\Delta W_{\rm in}\leq 0$. As a result, these represent
configurations that are either marginally stable (i.e.  $\Delta W_{\rm
in} = 0$) or even {\it stable} (i.e. $\Delta W_{\rm in} < 0$) with
respect to the runaway instability. Since these configurations cannot
develop the instability on a finite timescale, we have introduced
parametrized perturbations that would induce a small outflow through the
cusp. More specifically, we have modified the stationary equilibrium
configuration discussed in Section \ref{analytic} with a small radial
velocity which we have expressed in terms of the radial inflow velocity
characterising a relativistic spherically symmetric accretion flow onto a
Schwarzschild black hole, i.e. the Michel solution~\citep{michel:72}.
Using $\eta$ to parametrize the strength of the perturbation, we have
specified the initial radial (covariant) component of the three-velocity
as
\begin{equation}
\label{vpert}
v_r = \eta (v_{r})_{_{\rm Michel}} \ .
\end{equation}

	We regard this choice of initial data as a more realistic one for
at least two reasons. Firstly, in this way only a small region of the
fluid configuration, (i.e. the one located near the cusp) is effectively
out of equilibrium since $v^r$ falls for large radii. Secondly and more
important, an initial configuration of this type is much closer to the
one that might be produced in Nature. We recall, in fact, that tori of
the type considered here are expected to form in a number of different
events such as the collapse of supermassive neutron
stars~\citep{vietri_e_stella:98}, or the iron-core collapse of a massive
stars~\citep{macfayden:99}. Other scenarios for the
genesis of these objects
involve the coalescence of a binary system, either
consisting of two neutron stars (especially if they have unequal masses,
Shibata 2002) or consisting of a black hole and of a neutron star which
is then disrupted by the intense tidal field~\citep{lee:99a, lee:99b,
lee:00}. In all of these catastrophic events, the newly formed torus will
be initially highly perturbed and is expected to maintain also a radial
velocity in addition to the orbital one. In recent Newtonian simulations
performed by \citet{ruffert:99} the torus resulting from the dynamical
merging of two neutron stars was observed to oscillate and accrete onto
the newly formed black hole. The average inflow velocity in the central
region of the newly formed torus was measured to be $\sim 3\times
10^{-3}$, whereas at very small distances from the black hole the fluid
was infalling much more rapidly. To be consistent with the estimates
provided by \citet{ruffert:99}, we have chosen the parameter $\eta$ in
the range $[0.001, 0.06]$, corresponding to an average inward radial
velocity in the range $[0.0002, 0.04]$, respectively. However,
simulations with values as small as $\eta=0$ and as large as $\eta =
0.17$ have also been performed, but these have not introduced
qualitatively new features. It should also be noted that because the
orbital velocities are at least one order of magnitude larger than the
radial ones induced through the perturbations, the contribution of the
latter to the kinetic energy budget is rather small even when large
values of $\eta$ are considered.

	An aspect of our initial models worth underlining is that while
in principle the mass flux should be zero when $\Delta W_{\rm in} = 0$, 
this is never the case numerically. The unavoidable mass flux induced at 
the cusp can be made arbitrarily small after a suitable choice of the 
mass of the torus and of the strength of perturbation. This represents 
an important possibility because in the case of very small rest-mass 
accretion rates, the variations in the spacetime metric can be neglected 
and we can therefore investigate the response of the toroidal neutron 
stars in the absence of metric variations. We will refer to this regime 
as the one with a {\it ``fixed''} spacetime to distinguish it from the 
{\it ``dynamical''} spacetime regime, in which the accretion rate is not
negligible and metric functions need to be updated following the procedure 
of Section~\ref{ste}.

	With the choice of initial conditions discussed above, we have
evolved a large number of models covering only part of the relevant
parameter space. The properties of the different models are summarized in
Table~\ref{tab1}, which contains the ratio between the mass of the
toroidal star and of the black hole, the polytropic constant, the
specific angular momentum, as well as all the relevant radii of the
tori. Each of the models in Table~\ref{tab1} has been simulated for at
least four different values of the parameter $\eta$, both on a fixed and
on a dynamical spacetime. In all of the simulations we have kept fixed:
the adiabatic index (taken to be that of a degenerate relativistic
electron gas) $\gamma=4/3$, and the initial black hole mass, which we
have chosen to be $M = 2.5M_{\odot}$ for comparison with the results of
\citet{font:02}.

\section{Numerical results}
\label{results}

	In what follows we will discuss in detail the dynamics of the
perturbed toroidal neutron stars summarized in Table~\ref{tab1}. In
particular, we will first report the results about the runaway
instability and subsequently will discuss the long-term dynamics of the
toroidal neutron stars in response to perturbations.

\subsection{Dynamical spacetime: the runaway instability}
\label{runaway}

	Since the code used in our simulations is similar but distinct
from the one employed by \citet{font:02}, we have first tested it against
the results published by these authors. More precisely, we have
considered out-of-equilibrium initial conditions ($\Delta W_{\rm in}>0$),
evolving this configuration using the set of equations (\ref{fcf}) and
the metric update (\ref{met_up}). The results obtained agree (with
differences less than one percent) with those by \citet{font:02},
indicating that with this choice of initial conditions the runaway
instability develops rapidly, on timescales that are progressively
smaller as the mass ratio $M_{\rm t}/M$ and the initial potential jump,
$\Delta W_{\rm in}$, are increased. The occurrence of the instability is
signalled by the exponential growth of the rest-mass accretion rate which
rapidly reaches super-Eddington values (cf. Figure 8 of Font \& Daigne,
2002a). It should be noted that while the simulated accretion rates are
many orders of magnitude larger than the Eddington limit (this is $\sim
1.2\times 10^{-16}\ \mathrm{M_{\odot}/ s}$ for the black hole considered
here), these mass fluxes are also the ones required to account for the
large energetic release observed in $\gamma$-ray bursts.

\begin{figure}
\centerline{
\psfig{file=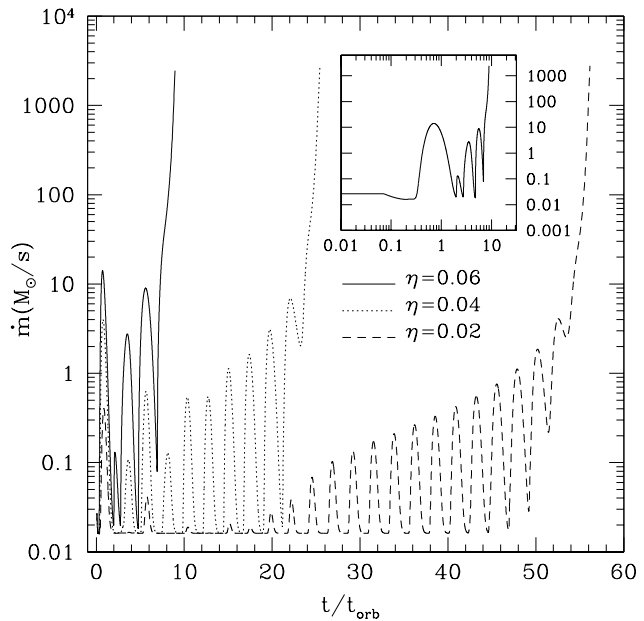,angle=0,width=8.5cm}
        }
\caption{
\label{fig3}
Evolution of the rest-mass accretion rate for model (a) and for different
values of $\eta$ when the spacetime is allowed to vary. The time is
expressed in terms of the timescale for the centre of the toroidal
neutron star to perform an orbit. The inset focuses on the case in which
$\eta=0.06$, showing the results on a logarithmic scale.}
\end{figure}

	After this validating test we have investigated the onset and
development of the runaway instability using the initial conditions
discussed in Section~\ref{initialdata}. Shown in Figure~\ref{fig3} is the
evolution of the rest-mass accretion rate for model $\rm (a)$ (see
Table~\ref{tab1}) and with three different values of initial velocity
perturbation, $\eta$. The time is expressed in terms of the orbital
period $t_{\rm orb} \equiv 2\pi/\Omega_{\rm centre}$ of the centre of the
toroidal neutron star and is reported in Table~\ref{tab1} for the
different models considered. Note that the minimum rest-mass accretion
rate in Figure~\ref{fig3} is never zero but $\sim 0.02 M_{\odot}/{\rm
s}$, even initially. This is just the cumulative result of the tenuous
ambient atmosphere, which is always producing a tiny and constant in time
mass overflow at the cusp, coupled with the use of very high density
matter which amplifies the effect.

	The behaviour of the mass flux reported in Figure~\ref{fig3} 
incorporates two important features. Firstly, it shows that the runaway 
instability does occur also with this choice of initial data and that 
the growth-rate is shorter for larger initial velocity perturbations 
(i.e. for larger values of $\eta$). As mentioned in the Introduction, 
this is an important point confirming that Roche lobe overflowing is 
not a necessary condition for the development of the instability, at 
least in constant specific angular momentum tori whose self-gravity 
is not considered. The inset of Figure~\ref{fig3} shows the simulation 
with $\eta=0.06$ on a logarithmic time scale, and should be considered 
the evolution that more than any other of our sample resembles the one 
observed for Roche lobe overflowing tori by \citet{font:02}.

	The behaviour observed in Figure~\ref{fig3} with different values
of the perturbation velocity has a simple interpretation. In order to
trigger the instability, in fact, a certain fractional change in the mass
of the black hole (and therefore in the spacetime curvature) needs to be
reached. If the initial perturbation is large, a considerable amount of
matter is accreted onto the black hole already after the first
oscillation in the accretion rate and the instability is therefore able
to develop very rapidly. If, on the other hand, the strength of the
perturbation is small, much less mass will be accreted during each
oscillation and many more will be needed to produce the fractional change 
in the black hole mass that will accelerate the development of the 
instability. In this case, the instability will develop on larger 
timescales, which do not scale linearly with $\eta$.

\begin{figure}
\centerline{ \psfig{file=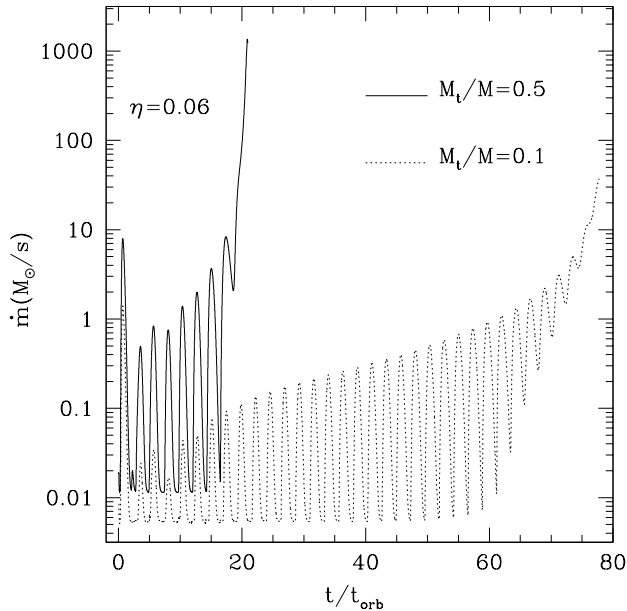,angle=0,width=8.5cm} }
\caption{
\label{fig7}
Rest-mass accretion rate for two different mass ratios in a dynamical
spacetime evolution. The solid and dotted lines correspond to models (b)
and (c), respectively, and a perturbation with $\eta=0.06$ was used.}
\end{figure}

	The second novel feature to notice in Figure~\ref{fig3} is that,
after the system has relaxed from the initial conditions (i.e. at about
$t/t_{\rm orb} \simeq 2$) the secular growth in the rest-mass accretion
rate is accompanied by an oscillatory behaviour with increasing
amplitude. These oscillations are less regular in the case of the high
amplitude perturbation (i.e. for $\eta=0.06$) but are much more regular
as the strength of the perturbation is gradually reduced. It is important
to notice that when very low amplitude perturbations are used, the amount
of accreted matter is so small that new features can be revealed. In the
case of the simulation for $\eta = 0.02$ in Figure~\ref{fig3} (dashed
line), for instance, it is possible to distinguish at least three
different stages. Most notably, an {\it initial} stage for $t/t_{\rm orb}
\lesssim 18$, during which the rest-mass accretion rate is very small and
does not manifest a regular oscillatory behaviour.  Despite the apparent
quiescence, and as it will become clearer in the following Section,
during this stage the toroidal neutron star is not at all static and
other hydrodynamical quantities manifest a different behaviour. This
stage is then followed by an {\it intermediate} stage for $18 \lesssim\;
t/t_{\rm orb}\; \lesssim 45$, during which the rest-mass accretion rate
shows the secular growth already observed for higher amplitudes
perturbations. While this happens, the toroidal neutron star periodically
enters short phases during which essentially no accretion is present
(i.e. when the accretion rate reaches its minimum).  Eventually, a third
{\it final} stage sets in for $t/t_{\rm orb} \gtrsim 45$, during which
the instability starts to develop more clearly. This can be deduced from
the fact that during this phase the accretion rate does not reach the
floor as it did in the previous intermediate stage and the oscillations
have progressively smaller amplitude while the accretion rate grows
exponentially.

	It is worth pointing out that during each of these oscillations a
considerable amount of matter falls towards the black hole and in a
realistic scenario it is reasonable to expect that before this matter
reaches the event horizon it will loose part of its potential binding
energy by increasing its temperature and by emitting electromagnetic
radiation. In view of this, it is plausible to expect that the
quasi-periodic accretion measured during our simulations could also be
observed in the form of a quasi-periodic X-ray luminosity, as it is
indeed the case in the quasi-periodic oscillations (QPO's) observed in
the X-ray luminosity of Low Mass X-ray Binaries
(LMXB's)~\citep{van_der_klis:00}. While the connection between the two
effects is very attractive, it should be remarked that the discs which
are believed to be behind the quasi-periodic X-ray luminosity in LMXB's
have much smaller rest-mass densities. Therefore, more detailed
calculations need to be done before any firm conclusion on the connection
between the two phenomenologies can be drawn.

\begin{figure}
\centerline{
\psfig{file=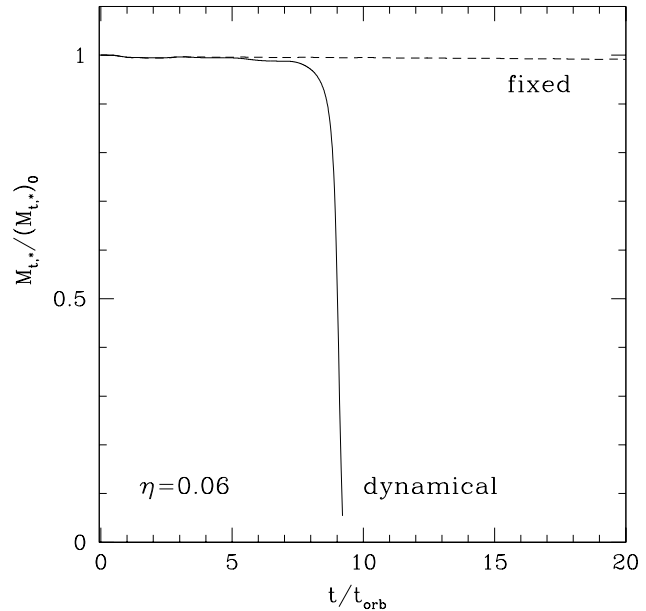,angle=0,width=8.5cm}
        }
\caption{
\label{fig15}
Evolution of the rest-mass of the toroidal neutron star (normalized to
its initial value) in a dynamical (continuous line) and in a fixed
(dashed line) spacetime simulation. The calculations refer to model (a)
with an initial perturbation $\eta=0.06$.  }
\end{figure}

	All of the calculations shown in Figure~\ref{fig3} terminate when
the accretion rate has reached a maximum value of $\sim 3\times 10^3
M_{\odot}/{\rm s}$ and the rest-mass of the torus has become only a few
percent of the initial one (cf. the solid line of
Figure~\ref{fig15}). During these very final stages of the instability
the calculations become very difficult because of the exponential changes
in the hydrodynamical quantities and Courant factors as small as $0.01$
are needed to prevent the code from crashing. Soon after the accretion
rate has reached its maximum, it drops rapidly to very small values as a
result of the almost complete disappearance of the torus (this final part
of the evolution is not reported in Figure~\ref{fig3}).

	As mentioned above, the growth-rate of the runaway instability
depends on the efficiency of the mass accretion process and on reaching a
certain fractional change in the black hole mass. How rapidly this change
takes place depends both on the strength of the initial velocity
perturbation (as shown in Figure~\ref{fig3}) but also on the density of
the accreted matter. To confirm this, we have performed simulations for
the same initial perturbation but with different mass ratios $M_{\rm
d}/M$. The results of these simulations are summarised in
Figure~\ref{fig7} which shows the behaviour of the rest-mass accretion
rate for models (b) and (c) in Table~\ref{tab1}. These models, we recall,
have the same properties of model (a) but have been constructed using
larger values of the polytropic constant $\kappa$ [cf. Eq.
(\ref{density})]. As expected, lower density toroidal neutron stars have
proportionally smaller accretion rates (note that the floors in the mass
flux ${\dot m}$ are different) and need longer timescales for the onset
and development of the instability.

	Two interesting aspects of Figure~\ref{fig7} need to be pointed
out. Firstly, the timescale for the instability to set in has an almost
linear dependence on the mass ratio $M_{\rm t}/M$. This is an important
detail as it reveals that what could be considered a ``realistic''
toroidal neutron star on the basis of the numerical simulations performed
so far~\citep{shibata:00, ruffert:01}, i.e. one with $M_{\rm t}/M \simeq
0.1$ and with a level of perturbations of the order $\eta \simeq 0.04$,
has a lifetime of roughly 0.2 s, if unstable to the runaway
instability. (cf. also with Figure~\ref{fig3}). Secondly, when the mass
accretion rate is generically low, the amount of matter accreted can be
very small even over several tens of dynamical timescales. When this is
the case, and as mentioned in Section~\ref{iap}, the spacetime can be
held fixed and the numerical calculations simplified. More importantly,
this regime provides the possibility of distinguishing the dynamical
response of the toroidal neutron star to perturbations from the
development of the instability. We will exploit this possibility in the
following Section which focuses on the investigation of the oscillation
properties of toroiodal neutron stars.

\subsection{Fixed spacetime: quasi-periodic oscillations}
\label{quasi}

\begin{figure}
\centerline{
\psfig{file=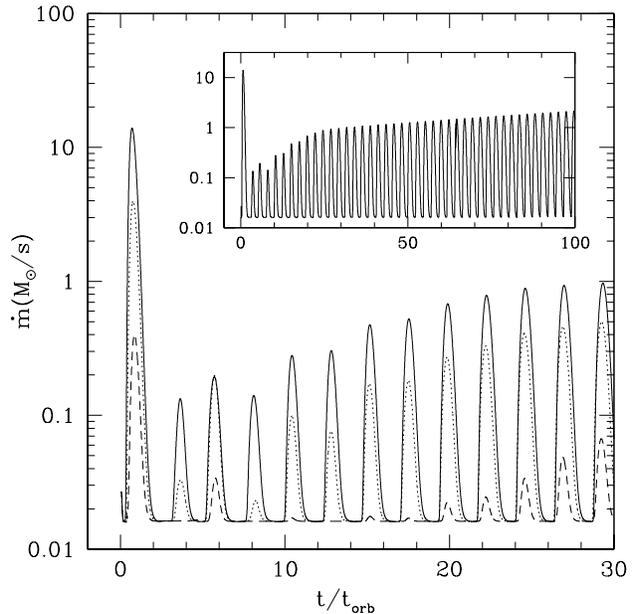,angle=0,width=8.5cm}
        }
\caption{Rest-mass accretion rate in a fixed spacetime evolution. The
solid, dotted and dashed lines correspond to $\eta=0.06, 0.04$ and
$0.02$, respectively. The inset shows the case $\eta=0.06$ during the
entire simulation (100 orbital periods).
\label{fig4}}
\end{figure}

	The dynamical response of the toroidal neutron star to
perturbations provides information about the basic properties of this
object in a strong gravitational field. While the details of these
properties depend on the details of the gravitational field the torus is
experiencing, we expect some features to be generic and therefore to be
present also in those circumstances in which the runaway instability is
suppressed. To prevent the runaway instability from hiding the
quasi-periodic response of the torus to perturbations, we have simply
suppressed the instability by maintaining the spacetime fixed.

\begin{figure}
\centerline{
\psfig{file=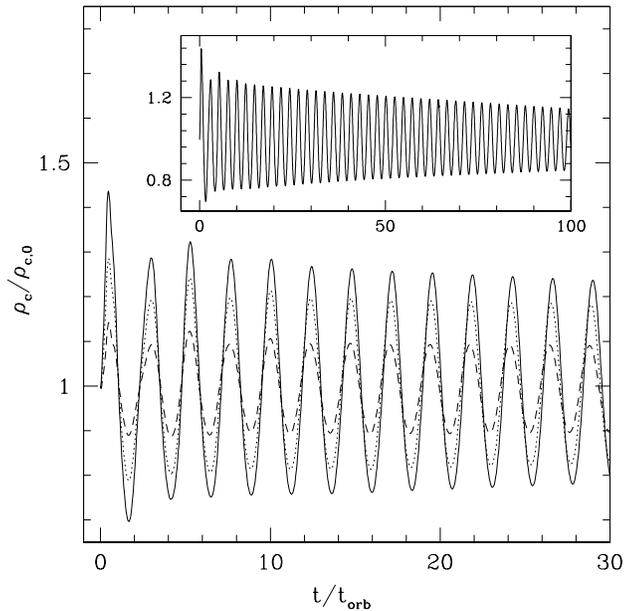,angle=0,width=8.5cm}
        }
\caption{
\label{fig5}
Evolution of the central rest-mass density normalized to its initial
value. As in Figure~\ref{fig4}, the solid, dotted and dashed lines
correspond to $\eta=0.06, 0.04$ and $0.02$, respectively. The inset shows 
the case $\eta=0.06$ during the entire simulation (100 orbital periods).}
\end{figure}

	The most apparent consequence of this choice is that the torus is
no longer dramatically accreted onto the black hole but remains at almost
constant rest-mass for arbitrarily long times. This is shown in
Figure~\ref{fig15} which displays the evolution of the rest-mass of the
torus (normalized to its initial value) for both a dynamical (dashed line)
and a fixed background spacetime, respectively. The solid line, in
particular, refers to the model shown in Figure~\ref{fig3} with the same
line style. Note that when the instability is fully developed the
toroidal neutron star has almost completely disappeared into the black
hole, the remaining mass being just $\sim 4\%$ of the initial one. Note
also that the rest-mass evolution in a fixed spacetime is not exactly
constant but shows a secular decay as a result of the small amounts of
matter that are quasi-periodically accreted onto the black hole
(cf. Figure~\ref{fig4}).

	Figure~\ref{fig4} is the equivalent of Figure~\ref{fig3}, showing
the quasi-periodic accretion rate during the first 30 orbital periods for
a simulation in which the spacetime is held fixed. It is apparent that
both evolutions have a qualitatively similar behaviour: after the
toroidal neutron star has relaxed at $t/t_{\rm orb} \simeq 2$, it starts
accreting matter onto the black hole at quasi-periodic intervals which do
not depend on the strength of the perturbation. The amplitude of the mass
accretion rate, on the other hand, does depend on the value of $\eta$,
producing larger amounts of accreted matter with increasingly larger
values of $\eta$. (Note that when $\eta=0.02$, the mass accretion rate
seems to drop to an almost constant value for $6 \lesssim t/t_{\rm orb}
\lesssim 15$. This is just the consequence of the logarithmic scale used
and, as shown in Figure~\ref{fig5}, it is indeed possible to observe a
periodicity also during this time interval.)

	Since the curves in Figure~\ref{fig4} are the result of numerical
calculations in which each period of oscillation requires several tens of
thousands timesteps, the ability to reproduce this periodicity with such
high precision is the result of the use of the accurate HRSC methods. In
addition to this, the periodic behaviour does not seem to be altered even
when observed over 100 orbital timescales as shown in the inset of
Figure~\ref{fig4}, although some secular features appear. Most notably,
the mass accretion rate oscillates around values that are increasingly
larger. This is due to the fact that as the accretion proceeds, matter of
increasingly larger rest-mass density reaches the cusp (the low-density,
outer regions of the toroidal neutron star have already been accreted)
and this therefore produces a small secular growth in the amplitude of
the mass flux.

	The mass accretion rate is not the only quantity showing a
periodic behaviour and indeed all of the fluid variables can be shown to
oscillate periodically. In Figure~\ref{fig5} we show the time evolution,
over 30 orbital periods, of the central rest-mass density of the toroidal
neutron star normalized to its initial value. The small inset shows the
evolution of the same quantity but for a much longer timescale and offers
convincing evidence of the regularity of the oscillation. The inset
should be contrasted with the evolution of the rest-mass density in a
simulation with a dynamical spacetime (not shown here). In that case, in
fact, the oscillations in the rest-mass density do not remain (roughly)
constant but grow exponentially as the runaway instability develops.

\begin{figure*}
\begin{center}
\hspace{0.001cm}
\psfig{file=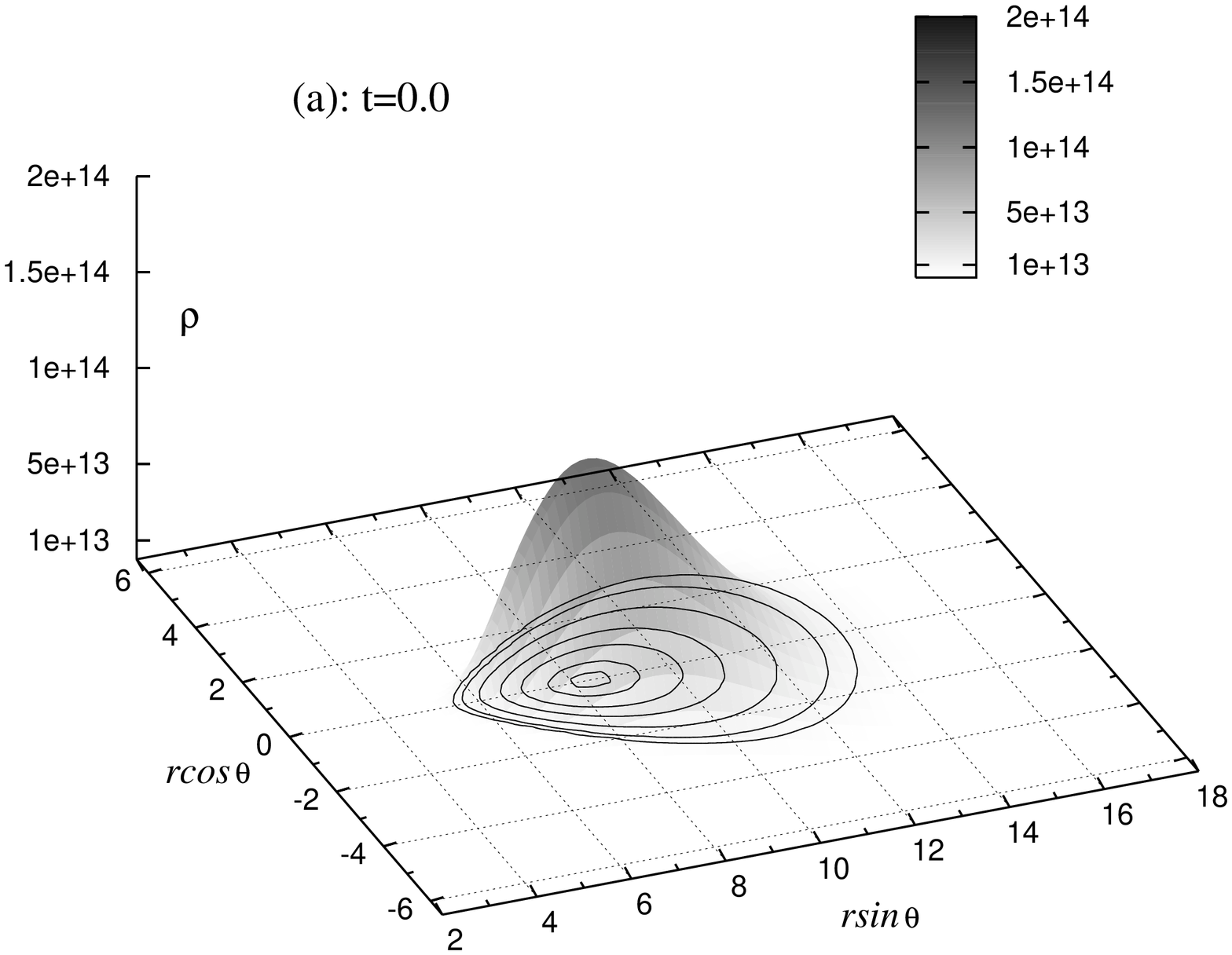,angle=-90,width=9.5cm,height=8.5cm}
\hspace{-1.50cm}
\psfig{file=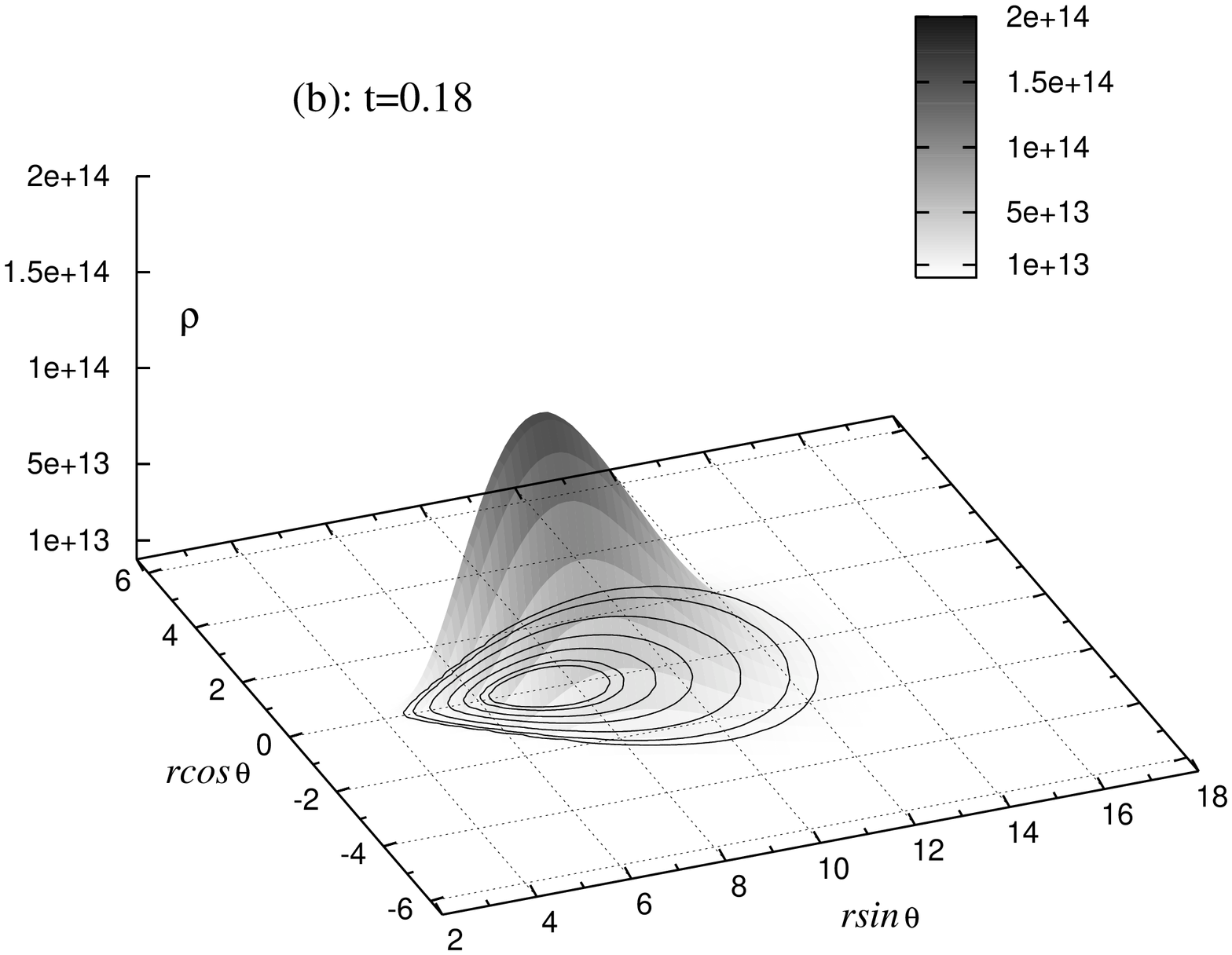,angle=-90,width=9.5cm,height=8.5cm}
\vspace{0.5cm}
\hspace{0.001cm}
\psfig{file=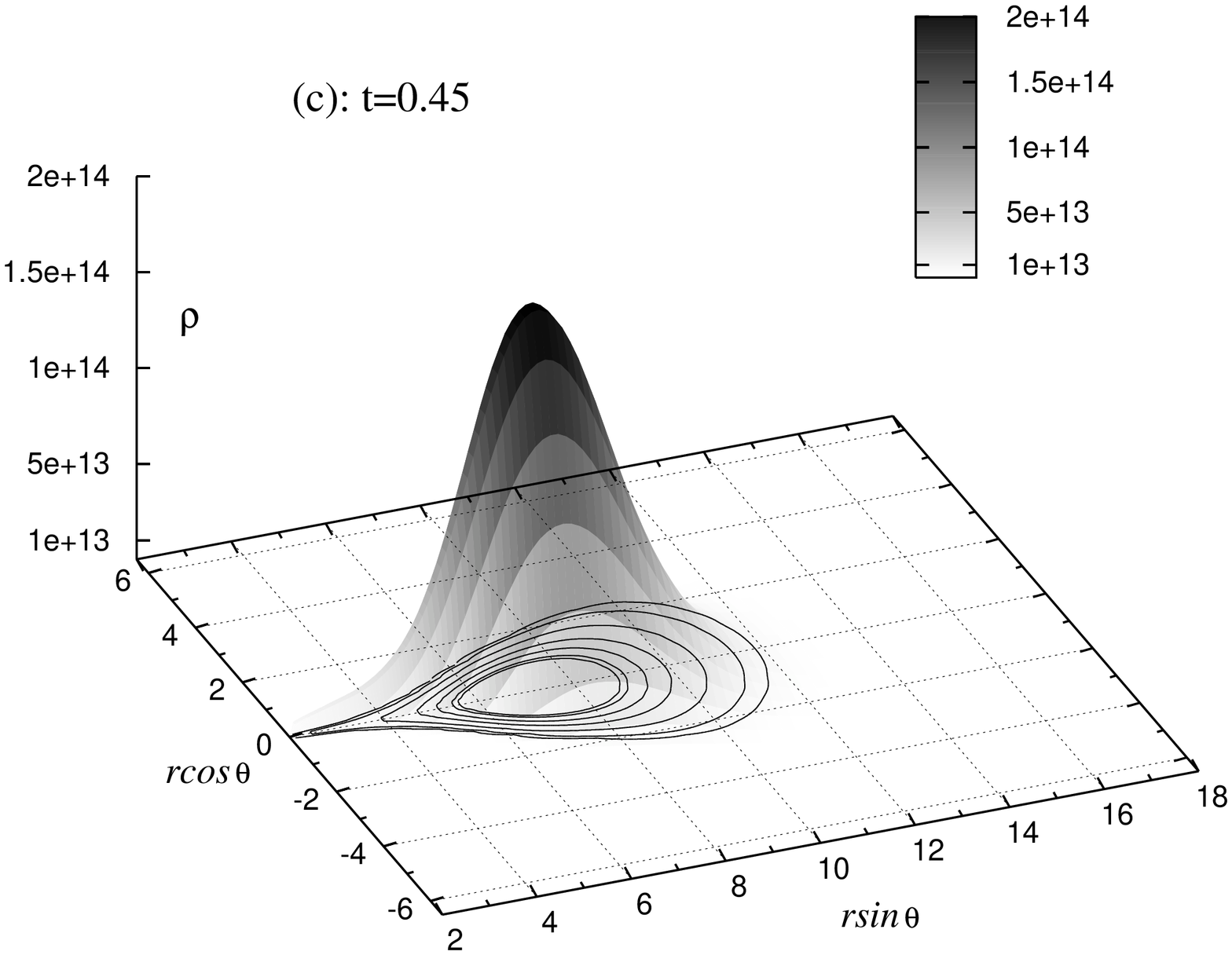,angle=-90,width=9.5cm,height=8.5cm}
\hspace{-1.50cm}
\psfig{file=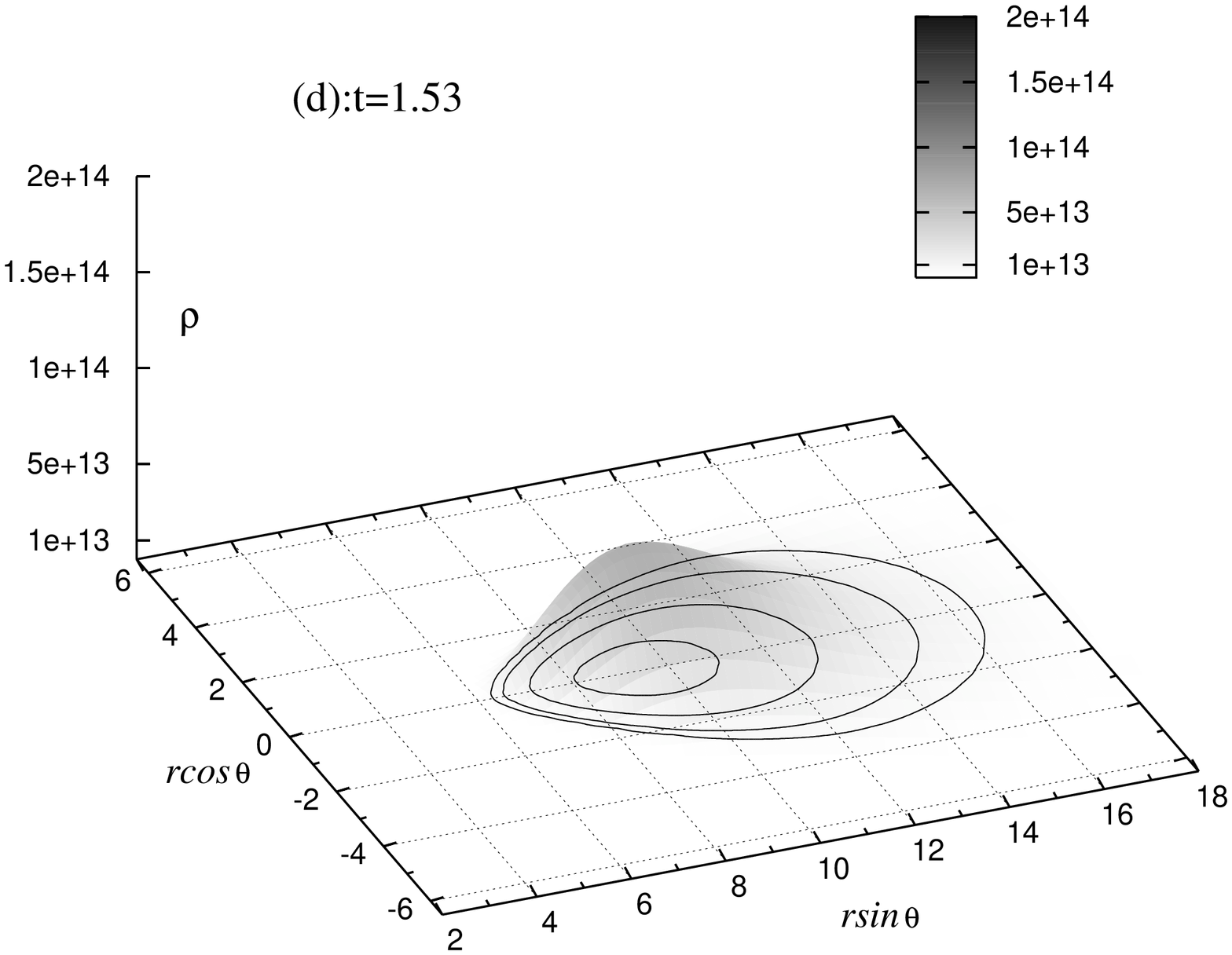,angle=-90,width=9.5cm,height=8.5cm}
\end{center}
\vspace*{-0.7cm}
\caption{Rest-mass density distributions of a perturbed toroidal neutron
star at different times. The sequence illustrates the periodic behaviour
during the first cycle of an oscillation excited with $\eta=0.06$. The
iso-density contours plotted on the $(r\sin\theta,r\cos\theta)$ planes of
the different panels can be used to trace the motion of the matter and
are the same for all of the panels (in particular they refer to $\rho=
5.0 \times 10^{\;12}, 1.\times 10^{13}, 2.5\times 10^{13}, 5.0\times
10^{13}, 7.5\times 10^{13}, 1.0\times 10^{14}$, and $1.11\times 10^{14}\
{\rm g/cm^3}$, respectively). The times are expressed in terms of the
orbital timescale, the spatial coordinates are in terms of the
gravitational radius, and the rest-mass density in ${\rm g/cm^3}$. See
main text for a comment on the sequence.}
\label{fig8}
\end{figure*}

	The periodic behaviour shown in Figure~\ref{fig5} has a simple
interpretation: as a result of the initial perturbation, the toroidal
neutron star acquires a linear momentum in the radial direction pushing
it towards the black hole. When this happens, the pressure gradients
become stronger to counteract the steeper gravitational potential
experienced as the torus moves inward, thus increasing the central density
and eventually pushing the torus back to its original position. This is
illustrated in more detail in the different panels of Figure~\ref{fig8},
which show the rest-mass distribution at various times during one
oscillation.

	More precisely, the sequence in Figure~\ref{fig8} shows that once
the unperturbed toroidal neutron star [whose initial rest-mass density
distribution is shown in the panel (a) of Figure~\ref{fig8}] is subject
to a radial velocity perturbation, it will start moving towards the black
hole [panel (b)]. The existence of a potential barrier at $r \simeq 4.5$,
however, causes a compression of the matter that is approaching the black
hole [panel (c)], giving rise to the first peak in the central density
visible in Figure~\ref{fig5} at $t/t_{\rm orb} \simeq 0.45$. Because the
initial configuration is just marginally stable, a small fraction of the
matter in the torus will be pushed over the maximum of the potential
barrier and generate the first maximum in the rest-mass accretion rate
reported in Figure~\ref{fig4}. The presence of a net mass flux onto the
black hole can directly be appreciated through iso-density contours shown
in the $(r\sin\theta, r\cos\theta)$ planes of Figure~\ref{fig8}. Most
notably, the lower density contours of panel (c) are closed on the event
horizon and indicate therefore the presence of a thin channel of
accreting matter that is linking the toroidal neutron with the black
hole.  (Because of this correlation between the rest-mass density and the
accretion rate, the peaks in Figure~\ref{fig5} are slightly advanced in
time with respect to the corresponding peaks in Figure~\ref{fig4}.) As
the compression increases, the pressure gradients become sufficiently
strong to produce a restoring force on the toroidal neutron star which is
then pushed back, away from the black hole. The restoring effect is so
efficient that the torus overshoots the original position [panel (a)] and
moves outwards to larger radii [panel (d)]. When this happens, the
central density decreases and the mass accretion rate drops to its floor
value; both of these effects are reflected in the first minima of
Figures~\ref{fig4} and \ref{fig5}.

	The dynamics of this process can also be followed by monitoring
the total energy of a fluid element at the edge of the torus, $(u_t)_{\rm
in}$. If, at a given time, this quantity becomes larger than the
potential barrier at the cusp, $W_{\rm cusp}$, the corresponding fluid
element will have sufficient kinetic energy to overcome the barrier but
not sufficient angular momentum to sustain an orbital motion at the
smaller radius at which it has been displaced. As a result, it will be
forced to fall into the black hole, producing, after its free-fall time,
a peak in the mass accretion rate.

	Once triggered, the behaviour described above will repeat itself
with great regularity (cf. the small insets of Figures~\ref{fig4} and
~\ref{fig5}) and minimal numerical dissipation (which can be appreciated
through the very small decay in the amplitude over 100 dynamical
timescales) up until the numerical simulation is stopped or the toroidal
neutron star has been entirely accreted by the black hole. During these
oscillations, the pressure gradients act as a restoring force during the
periodic transformation of the excess kinetic energy (transferred with
the initial velocity perturbation) into potential gravitational energy
and viceversa. As we shall comment on later, this is a first clue about
the nature of these pulsations.

	It is worth noting that the introduction of a perturbation in the
radial velocity will no longer guarantee that the specific angular
momentum is conserved during the time evolution. This can be most easily
seen if we rewrite the $\phi$-component of the Euler equations in the
form [cf. Eq. (24) of \citet{hawley:84a}]
\begin{equation}
\label{dldt}
\partial_t \ell = \frac{1}{\rho h \Gamma^2\sqrt{-g}}
\partial_i \left[\sqrt{-g} \rho h \Gamma^2 v_{\phi} v^i\right] \ .
\end{equation}
For an unperturbed toroidal neutron star, $v^{r} = v^{\theta} = 0 =
\partial_{\phi}$ and the right-hand-side of Eq. (\ref{dldt}) is therefore
zero, ensuring the conservation of the specific angular momentum.
Clearly, this is no longer true when a small negative radial velocity is
introduced and although small, this effect further favours the tiny spill
of matter through the edge of the torus.

	A final comment should be made about initial models consisting of
initially stable toroidal neutron stars $(\Delta W<0)$ and that are
therefore fully contained in barotropic surfaces smaller than their Roche
lobe. Also for these models we have performed a number of simulations
investigating their behaviour for different initial perturbations as well
as for different initial masses. Overall, the behaviour observed with
these initial data is qualitatively similar to the one already discussed
for marginally stable tori, i.e. also these models develop the runaway
instability or show a quasi-periodic behaviour depending on whether the
spacetime evolution is taken into account or not. The most significant
difference observed is that smaller rest-mass accretion rates are
generally produced for the same initial perturbation. This is simply due
to the fact that in stable models a larger potential barrier is present
at the cusp and is therefore increasingly more difficult for a fluid
element to reach the black hole as a result of the initial perturbation.

\begin{figure}
\centerline{
\psfig{file=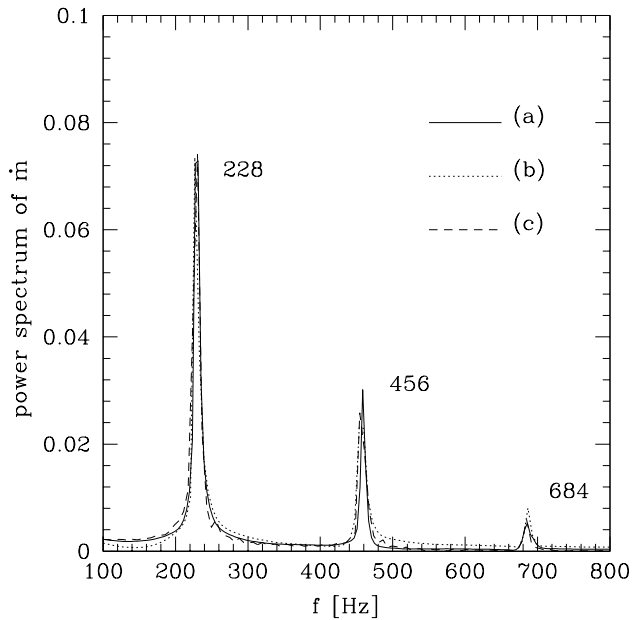,angle=0,width=8.5cm}
        }
\caption{Power spectrum of the mass accretion rate for models (a), (b),
and (c). The values on the vertical axis have been suitably normalized to
match the power in the fundamental mode and arbitrary units have been
used. }
\label{fig13}
\end{figure}

\subsubsection{Fourier analysis}
\label{fa}

	We have so far commented on the ``quasi-periodic'' behaviour of
the hydrodynamical variables in response to the initial perturbation but
we have not yet discussed how {\it periodic} is ``quasi-periodic''. We
have therefore calculated the Fourier transforms of the relevant fluid
quantities for a number of different models. As a good representative
case, we plot in Figure~\ref{fig13} the power spectrum of the rest-mass
accretion for models (a), (b) and (c) of Table~\ref{tab1}. The Fourier
transform has been calculated with data obtained with an $\eta=0.06$
perturbation and computed over a time interval going up to $t/t_{\rm orb}
\simeq 100$. Note that larger values of $\eta$ produce correspondingly
higher peaks in the power spectra, but the data in Figure~\ref{fig13} has
been suitably normalized to match the power in the first peak.

	There are two important features of Figure~\ref{fig13} that need
to be pointed out. The first one is that all of the three power spectra
shown consist of a fundamental frequency $f_0$ (228 Hz for the models
considered in Figure~\ref{fig13}) and a series of overtones (at 456 and
684 Hz, respectively) in a ratio which can be determined to be $1:2:3:
\ldots$ and to an accuracy of a few percent. (Note that overtones higher
than the second one have also been measured, although with much lower
power than the one found in the first three peaks.) Indeed, the presence
of at least three peaks can be detected also in the power spectra of
basically all of the fluid variables as well as in the overall
displacement of the toroidal neutron star during the
oscillations. Interestingly, the power spectra of some fluid variables,
such as that of the $L_2$ norm of the rest-mass density, seem to show
peaks with much lower power also at intermediate frequencies, in
particular at frequencies which are in a ratio $1 : 3/2 : 2 : 5/2 :
\ldots$ with respect to the fundamental one. Because the energy of these
peaks is very close to the background noise, it is not clear whether
these modes correspond to physical modes or are due to numerical
errors. The second important feature is that the peaks in the power
spectra of the three models in Figure~\ref{fig13} have all the same
frequencies, with differences below $0.1\%$. All of these properties are
clearly suggesting that the quasi-periodic response observed is the
consequence of some fundamental mode of oscillation of toroidal neutron
stars and that, as for isolated neutron stars, it is probably independent
even of the presence of a central black hole.

\begin{figure}
\centerline{
\psfig{file=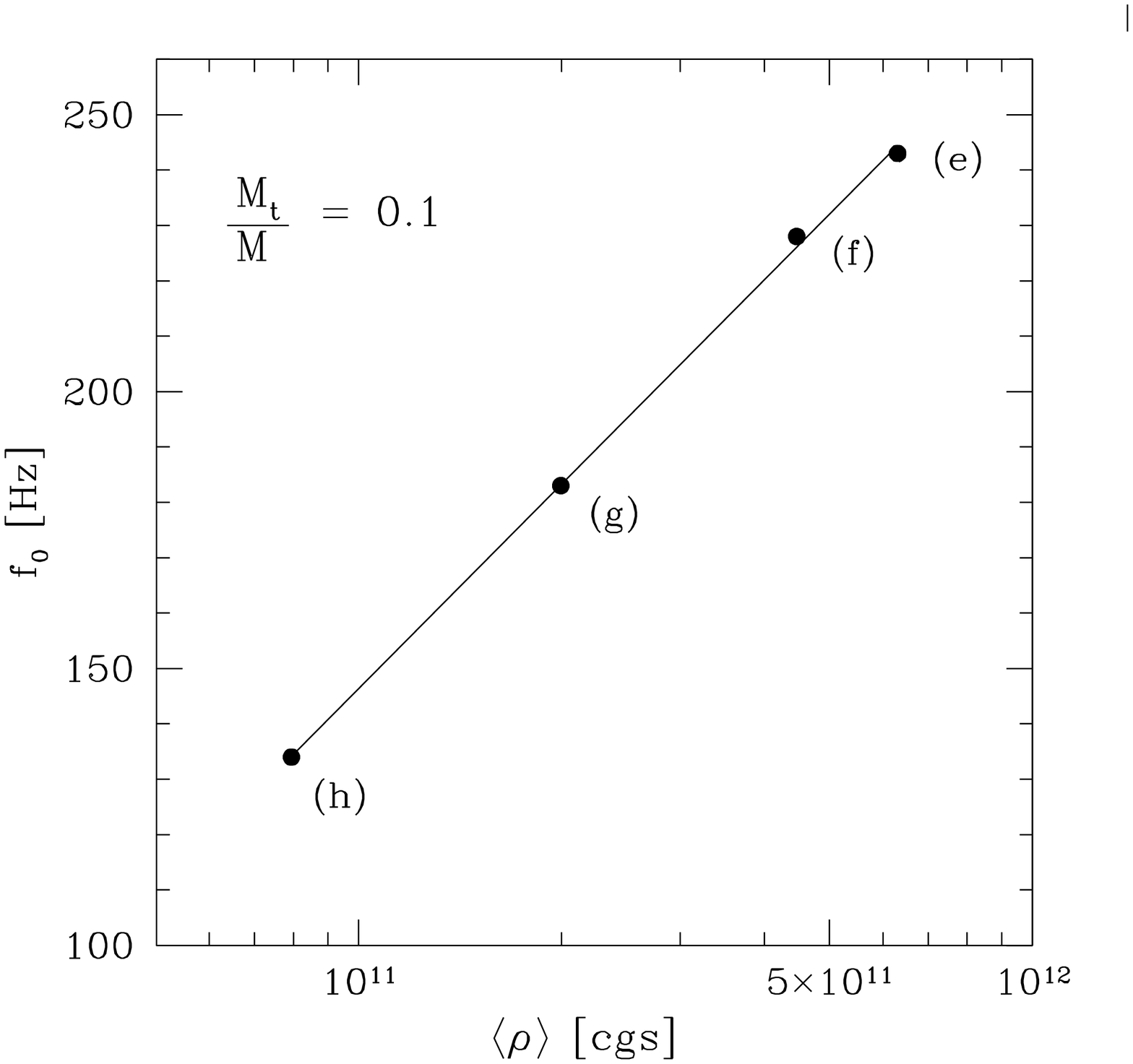,angle=0,width=8.5cm}
        }
\caption{Dependence of the fundamental frequency $f_0$ on the average
density in a sequence of models having the same mass ratio,
$M_{\rm{d}}/M=0.1$. The data refers to simulations with initial
perturbations of strength $\eta=0.06$. }
\label{fig2}
\end{figure}

	Insight on the nature of these modes can be gained if one
considers that all of the models reported in Figure~\ref{fig13} refer to
toroidal neutron stars with fixed spatial dimensions and specific angular
momentum, but with varying mass (cf. Table~\ref{tab1}).  This has
basically been obtained by suitably rescaling the polytropic constant in
the EOS. Another property shared by these models is that they all have
the same initial average sound speed. We recall that for the ideal fluid
configurations considered here, the local sound speed can be calculated
as $c_s=\sqrt{{\gamma(\gamma - 1)p}/[{(\gamma - 1)\rho+\gamma p}]}$,
which is effectively a constant for models with an initial density
distribution given by Eq. (\ref{density}). As a result, it is not
surprising that the peaks coincide in all of the models if the
oscillations discussed so far should be associated to the {\it p-modes}
(or acoustic modes) of the toroidal neutron star. However, proving that
the periodic oscillations observed in our simulations can be interpreted
as $p$-modes and validating that the presence of modes at intermediate
frequencies is not a numerical artifact requires a detailed perturbative
analysis and is beyond the scope of this paper. Work is now in progress
and preliminary results in this direction seem to confirm the hypothesis
that these oscillations represent the vibrational modes of relativistic
toroidal neutron stars having time-varying pressure gradients as the
restoring force~\citep{ryz:02}.

	Since the peaks in the power spectra seem related to the average
sound speed, the dependence of the fundamental frequency on the
properties of the toroidal neutron star needs to be investigated along
sequences different from the one consider in Figure~\ref{fig13}. As an
example, we have reported in Figure~\ref{fig2} the fundamental frequency
$f_0$ as a function of the average rest-mass density inside the
torus. The data refer to the evolution of models (e)--(h) in
Table~\ref{tab1}, that have different dimensions, specific angular
momenta and polytropic constants, but all have the the same mass ratio
$M_{\rm t}/M$ (cf. Table~\ref{tab1}). Figure~\ref{fig2} shows that an
evident correlation exists between the fundamental frequency and the
logarithm of the average rest-mass density. A fit to the data indicates
that this correlation is (cf. straight line in Figure~\ref{fig2})
\begin{equation}
\label{fit1}
f_0 = (122.55 \log{\langle{\rho}\rangle} - 1201.67)~{\rm Hz} \ ,
\end{equation}
where $\langle{\rho}\rangle$ is expressed in cgs units. This expression
is important as it suggests that a systematic study of these oscillations
for different initial models of toroidal neutron stars is
possible. Furthermore, it represents a first step towards a relativistic
disc-seismology analysis for massive and vertically extended tori in
General Relativity, in analogy to the one extensively developed for
geometrically thin discs~\citep{kato:01, perez:97, silbergleit:01,
rodriguez:02}.

\subsubsection{Linear and nonlinear regimes}

\begin{figure}
\centerline{
\psfig{file=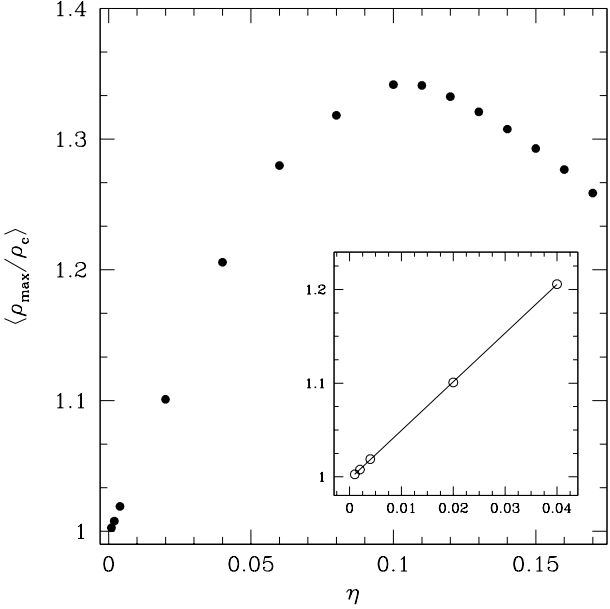,angle=0,width=8.5cm}
        }
\caption{Averaged maximum rest-mass density normalized to the central one
in the toroidal neutron star as a function of the perturbation
strength. The small inset shows a magnification of the behaviour in the
linear regime and the solid line shows a very good linear fit to the
data. The data refers to model (a) and has been averaged over 10 orbital
timescales.
\label{fig14}}
\end{figure}

	All of the quasi-periodic behaviour discussed so far is the
consequence of the finite size perturbations that have been introduced in
the initial configuration. Within this approach a {\it linear} regime is
expected in which the response of the toroidal neutron star is linearly
proportional to the perturbation introduced, and a {\it nonlinear} regime
when this ceases to be true. The strength of the perturbation which marks
the transition between the two regimes can be estimated from
Figure~\ref{fig14}, where we show the averaged maximum rest-mass density
normalized to the central one in the toroidal neutron star.

	A rapid look at Figure~\ref{fig14}, in fact, reveals the presence
of both the linear and nonlinear regimes with the first one being shown
magnified in the inset, and where the solid line shows the very good
linear fit to the data. The transition between the two regimes seems to
occur for $\eta \simeq 0.05$, with the nonlinear regime producing maximum
amplitudes that are $\sim 35\%$ larger than the initial one. A careful
analysis of the behaviour of the fluid variables shows that for
perturbations with strength $\eta \gtrsim 0.05$, some of the kinetic
energy which is confined to the lower order modes in the linear regime,
tends to be transferred also to higher order modes (As remarked above,
while in Figure~\ref{fig13} we have reported only the first three peaks,
the power spectra show the presence of higher order overtones, with up to
the seventh one being clearly visible). The nonlinear coupling among
different modes and the excitation of higher order overtones is often
encountered in Nature where it serves to redistribute the excess kinetic
energy before the production of shocks. In practice, the nonlinear
coupling deprives of energy the fundamental mode (which is the one
basically represented in Figure~\ref{fig14}) and is therefore responsible
for the decay of $\langle \rho_{\rm max}/\rho_{\rm c}\rangle$ for $\eta
\gtrsim 0.1$. Interestingly, when analysed in terms of the power spectra,
this effect shows a very distinctive behaviour. As the nonlinear
mode-mode coupling becomes effective, the amount of power in the
fundamental mode becomes increasingly smaller as the strength of the
perturbation is increased. At the same time, the conservation of energy
transfers power to the overtones, with the first ones reaching amplitudes
comparable to the fundamental one and with the high order ones becoming
more and more distinct from the background.

	Determining the transition to the nonlinear regime is important
to set an approximate upper limit on the amplitude of the oscillations
and, as we will discuss in the following Section, it is relevant
when estimating the emission of gravitational waves. It should also be
noted that in the parameter range for $\eta$ in which we have performed
our calculations (i.e. $\eta \in [0.001,0.12]$), the peak frequencies in
the power spectra have not shown to depend on the values used for $\eta$.
This is of course consistent with them being fundamental frequencies
(and overtones).

	A final comment should be made on the minimum value of the
perturbation parameter $\eta$ which is sufficient to produce the
quasi-periodic behaviour described in this Section. On the basis of the
continuum equations one expects that this minimum value is strictly
larger than zero. However, we have performed simulations of marginally
stable configurations $(\Delta W=0)$ with $\eta=0$ and observed much of
the phenomenology described above even if at rather minute amplitudes and
with a larger numerical noise. This result, which has been encountered
also in other accurate simulations of relativistic
stars~\citep{font-goodale:02}, is not entirely surprising and is simply
indicating that in the absence of prescribed perturbations, even the
small truncation error introduced in the construction of the initial
configuration is sufficient to excite the pulsations in these modes of
vibration.

\section{Gravitational wave emission}
\label{gwe}

	Despite our analysis has been restricted to axisymmetry, two
simple considerations suggest that strong quasi-periodic gravitational
waves should be expected together with the quasi-periodic accretion. The
first of these considerations is that these oscillating tori undergo
large and rapid variations of their mass quadrupole moment which, we
recall, can be calculated as
\begin{equation}
\label{mq}
I\equiv \int \rho \left(\frac{3}{2}z^2 - \frac{1}{2}\right)r^4 dr dz \ ,
\end{equation}
where $z\equiv \cos\theta$. It is therefore reasonable to expect that
gravitational waves, with a strain proportional to the second time
derivative of Eq.~(\ref{mq}), should be produced during such
oscillations\footnote{It is worth underlying that a deviation from
axisymmetry is only a sufficient condition for the emission of
gravitational waves and not, as sometimes stated~\citep{mineshige:02}, a
necessary condition.}.  The second consideration is suggested by
expression (\ref{mq}), which shows that a configuration with toroidal
topology and in which the rest-mass density has a maximum away from the
origin will naturally have a large mass quadrupole simply because of the
product $\rho r^4$ in the integrand of this equation. This should be
contrasted with what happens for stars with spherical topology and which
have instead the largest densities at the centre.  Both of these
arguments justify the intense gravitational radiation and the large
signal-to-noise ratio we have calculated using the Newtonian quadrupole
approximation. Before discussing this in detail, however, it is useful to
remind that since we are not solving the Einstein field equations, we are
unable to account for that part of the gravitational radiation that is
emitted by the black hole itself as a result of the quasi-periodic mass
accretion. For the same reason we cannot estimate the amount of
gravitational radiation that will be captured by the black hole and will
not reach null infinity.  Work is now in progress to calculate also this
part of the radiative field using an approach in which the fluid
evolution is used as a ``source'' for a perturbative form of the Einstein
equations.

	The presence of an azimuthal Killing vector has two important
consequences. Firstly, the gravitational waves produced by these
axisymmetric oscillating tori will carry away energy but not angular
momentum, which is a conserved quantity in this spacetime. This is to be
contrasted with the gravitational wave emission from non-axisymmetric
perturbations in a massive torus orbiting a black hole, whose strength
has been estimated in recent papers~\citep{van_putten:01a,
van_putten:01b, mineshige:02}. Secondly, the gravitational waves produced
will have a single polarization state (i.e. the ``plus'' one in our
coordinate system,~\citet{kochanek:90}), so that the transverse traceless
gravitational field is completely determined in terms of its only nonzero
transverse and traceless ({\it TT}) independent components,
$h^{TT}_{\theta\theta} = -h^{TT}_{\phi \phi}$~\citep{zwerger:97}.
Adopting then Newtonian quadrupole approximation, we can calculate the
gravitational waveform $h^{TT}(t)$ observed at a distance $R$ from the
source in terms of the quadrupole wave amplitude $A_{20}^{\rm E2} $ (see
also Zwerger \& M\"uller 1997)
\begin{equation}
\label{htt}
h^{TT}(t) = F_+ \left(\frac{1}{8}
	  \sqrt{\frac{15}{\pi}}\right) \frac{A_{20}^{\rm E2}(t-R)}{R} \ ,
\end{equation}
where $F_+=F_+(R,\theta,\phi)$ is the detector's beam pattern function
and depends on the orientation of the source with respect to the
observer.  As customary in these calculations, we will assume it to be
optimal, i.e. $F_+=1$. The wave amplitude $A_{20}^{\rm E2}$ in
Eq. (\ref{htt}) is simply the second time derivative of the mass
quadrupole moment and can effectively be calculated without taking time
derivatives, which are instead replaced by spatial derivatives after
exploiting the continuity and the Euler equations~\citep{finn:89,
blanchet:90, rezzolla:99} to give
\begin{eqnarray}
\label{stress}
&&A_{20}^{\rm E2} \equiv \frac{d^2 I}{dt^2} =
	k \!\int\!\! \rho \biggl[v_r v^r (3 z^2 \!-\! 1) \!+\!
	v_\theta v^\theta (2 \!-\! 3 z^2) \!-\!
	v_\phi v^\phi  \nonumber \\
	& & \qquad \qquad \qquad -6 z \sqrt{(v^r v_r)(v_\theta v^\theta)
	(1 \!-\! z^2)} \biggl. - r \frac{\partial \Phi}{\partial r}
	(3 z^2 \!-\! 1) \nonumber \\
	& & \qquad \qquad \qquad\, + 
	\, 3 z\frac{\partial \Phi}{\partial \theta}
	\sqrt{1 \!-\! z^2}\biggr] r^2 {\rm d}r {\rm d}z \ ,
\end{eqnarray}
where $ k = 16 \pi^{3/2} / \sqrt{15} $, and $\Phi$ is the gravitational
potential. Since Eq.~(\ref{stress}) is intrinsically Newtonian, it brings
up two subtle issues when evaluated within a relativistic context. These
basically have to do with the definition of the radial coordinate and
with the definition of the gravitational potential appearing in
Eq.~(\ref{stress}). We have here opted for a pragmatical approach and
treated $r$ as the Schwarzschild radial coordinate and computed the
gravitational potential in terms of the radial metric function as $\Phi =
(1 - g_{rr})/2$, which is correct to the first Post-Newtonian (PN) order.

	To validate the correct implementation of the integral in
Eq.~(\ref{stress}), we have considered a stationary torus in stable
equilibrium, (i.e. $\Delta W_{\rm in}<0$) and without any perturbation
besides the one introduced by the truncation error (i.e. $\eta=0$). Under
these circumstances, no gravitational radiation can be produced and the
terms in the square brackets of Eq.~(\ref{stress}) should therefore
compose to give zero identically. When computed numerically, we have
found that the sum of these terms is effectively very small and of the
order $\sim 10^{-2}$. This small residual in the integrand is due to
approximations mentioned above (i.e. the use of the Schwarzschild radial
coordinate and the first PN approximation to the gravitational potential)
and should be interpreted, in practice, as a consequence of the fact that
our tori are highly relativistic objects, whose equilibrium is not
exactly given by the balancing of the Newtonian terms in the square
brackets.  Because of this residual, however, the wave amplitude
$A_{20}^{\rm E2}$ will not average to zero over an oscillation but will
have a net offset. We account for this by removing the overall residual
in the evaluation of the wave amplitude (see also Dimmelmeier et al. 2002
for the discussion of an analogous technique).

\begin{figure}
\centerline{
\psfig{file=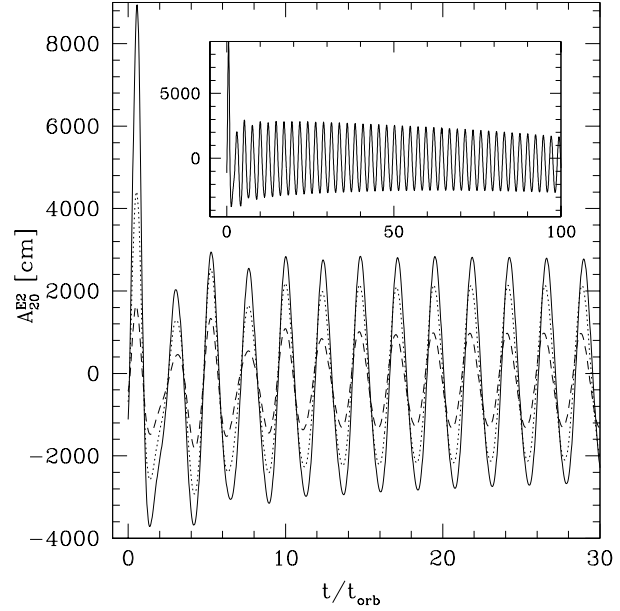,angle=0,width=8.5cm}
        }
\caption{
\label{fig6}
Time evolution of the second time derivative of the mass quadrupole,
computed for model (a). The solid, dotted and dashed lines correspond to
$\eta=0.06, 0.04, 0.02$, respectively. The inset corresponds to the case
$\eta=0.06$ and spans 100 orbital timescales.}
\end{figure}

	Figure~\ref{fig6} shows the time evolution of the wave amplitude
$A_{20}^{\rm E2}$ computed for model (a) via the expression
(\ref{stress}).  The different line types refer to $\eta=0.06, 0.04$ and
$0.02$, respectively.  The computed gravitational waveform exhibits the
same periodic behaviour discussed in the previous Sections for several
fluid variables and shows oscillations that are in phase with the ones
observed in the rest-mass density (cf. Figure~\ref{fig5}).  Furthermore,
and as one would expect, the wave amplitude scales with the strength of
the initial perturbation. Note that in the case of a simulation with a
dynamical spacetime (not shown in Figure~\ref{fig6}), the gravitational
waveform does not maintain a constant in time amplitude but, as the
runaway instability develops, the variations in $A_{20}^{\rm E2}$ become
increasingly large, with an exponential growth rate that matches the one
observed in the density evolution.

\begin{figure}
\centerline{
\psfig{file=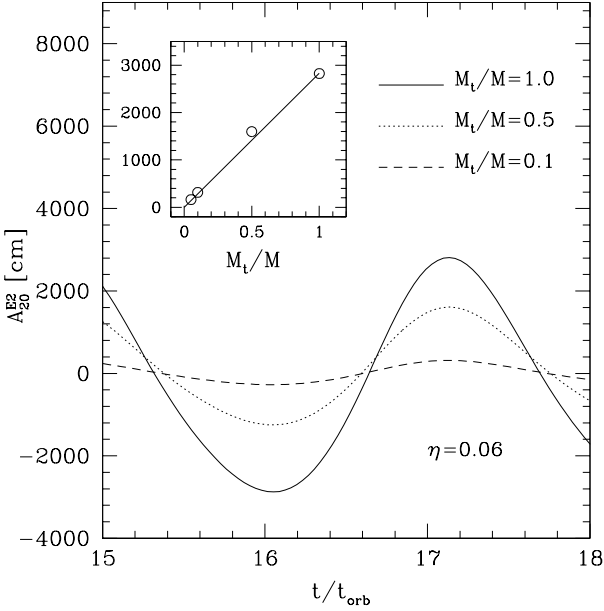,angle=0,width=8.5cm}
        }
\caption{Dependence of the wave amplitude $A^{\rm E2}_{\rm 20}$ on the
mass of the torus. The solid, dotted and dashed lines correspond to models
(a), (b) and (c), respectively. The data refers to an initial
perturbation of strength $\eta=0.06$ and the inset shows a linear fit
(solid straight line) of the data, including the one for model (d).
\label{fig9}
}
\end{figure}

	Since the mass quadrupole and its second time derivative are
linear in the rest-mass density [cf. Eq. (\ref{stress})] and the latter
exhibits a quasi-periodic behaviour upon perturbations, it is natural to
expect the same linear dependence to be present also in terms of the mass
ratio $M_{\rm t}/M$. To verify this we have computed $A_{20}^{\rm E2}$
with an initial perturbation $\eta=0.06$ for models (a), (b), and (c)
that, we recall, differ only for their mass $M_{\rm t}$. The results of
this analysis are reported in Fig.~\ref{fig9} which shows the behaviour
of $A_{20}^{\rm E2}$ over one representative period of oscillation and
for the four models. As expected, the scaling of the amplitude is linear
with $M_{\rm t}/M$ and this is more clearly shown in the inset where the
wave amplitude [including the data for model (d)] is fitted linearly with
$M_{\rm t}/M$.

	Using Eq.~(\ref{htt}), it is possible to derive a
phenomenological expression for the gravitational waveform that could be
expected as a result of the oscillations induced in the toroidal neutron
star. Restricting our analysis to the linear regime for which a simpler
scaling is possible, we can express the transverse traceless
gravitational wave amplitude for a source in the Galaxy in terms of the
relevant parameters in our problem
\begin{equation}
\label{h_phenom}
h^{TT} \simeq 2.2\times 10^{-21} \left(\frac{\eta}{0.04}\right)
	\left(\frac{M_{\rm t}}{0.1\ M_{2.5}}\right)
	\left(\frac{10\; {\rm Kpc}}{R}\right) \ , 
\end{equation}
where we have defined $M_{2.5}\equiv M/(2.5\ M_{\odot})$.

	Overall, expression~(\ref{h_phenom}) shows that, already in the
linear regime for the perturbations, a non-negligible gravitational wave
amplitude can be produced by an oscillating toroidal neutron star
orbiting around a black hole. This amplitude is indeed comparable with
the average gravitational wave amplitude computed in the case of core
collapse in a supernova explosion \citep{zwerger:97,dimmel:02} and can
become stronger for larger perturbations or masses in the torus.


	Of course, the large wave amplitudes suggested by expression
(\ref{h_phenom}) are referred to a galactic source and would become three
orders of magnitude smaller for a source located at the edge of the Virgo
cluster (i.e. at about 20 Mpc). What is important to bear in mind from
expression (\ref{h_phenom}) is that oscillating toroidal neutron stars
can be sources of gravitational waves {\it as strong or stronger} than a
core collapse and that could occur with {\it comparable} event
rates\footnote{Of course not all of the collapsing iron cores will produce a
black hole sourrounded by a massive torus. On the other hand, and as
discussed in Section~\ref{iap}, there is a multiplicity of scenarios in
which toroidal neutron stars could be produced.}. This notion then serves
as a useful normalization for estimating their relevance. It should also
be noted, however, that a strong gravitational wave signal is just a
necessary condition for the detectability of the gravitational wave
emission from toroidal neutron stars and that additional conditions, such
as a sufficiently high event rate or a good matching with the sensitivies
of the detectors, need to be met. In the following Section we will
discuss these issues in more detail and evaluate the detectability of
these potential new sources of gravitational waves.

\subsection{Detectability}

	To assess the detectability of toroidal neutron stars as sources
of gravitational radiation we have computed the {\it characteristic}
gravitational wave frequency and amplitude, as well as the corresponding
signal-to-noise ratio for the interferometric detectors that will soon be
operative. More specifically we have first computed the gravitational
waveform in the frequency domain as the Fourier transform of the
traceless transverse waveform in the time domain
\begin{equation}
\label{h_of_f}
{\tilde h}(f) \equiv \int_{-\infty}^{\infty}\, 
	e^{2\pi i f t}h^{TT}(t)\, dt \ ,
\end{equation}
where $h^{TT}(t)$ is calculated according to Eq.~(\ref{htt}) and where we
have considered the gravitational wave amplitude computed over a {\it
fixed} spacetime only. (Hereafter we will indicate $h^{TT}(t)$ simply as
$h(t)$.). While in principle the integral in Eq.~(\ref{h_of_f}) is over
an infinite time interval, in practice $h(t)$ is nonzero only over a
finite interval $\tau_{\rm life}$. If the initial model chosen for the
toroidal neutron star is not a stable one, this time interval is simply
set by the timescale over which the runaway instability develops. If, on
the other hand, the initial model is stable, the timescale over which a
gravitational wave signal is produced can be considerably longer and is
basically set by the time over which the star is able to survive, for
instance, against non-axisymmetric instabilities. Hereafter, as a
representative timescale for our ``realistic'' toroidal neutron star
model (cf. Section \ref{runaway}) we will assume $\tau_{\rm life} \simeq
0.2$ s, while longer/shorter timescales will be assumed for models with
smaller/larger initial perturbations (cf. Table~\ref{tab2}).

\begin{figure}
\centerline{
\psfig{file=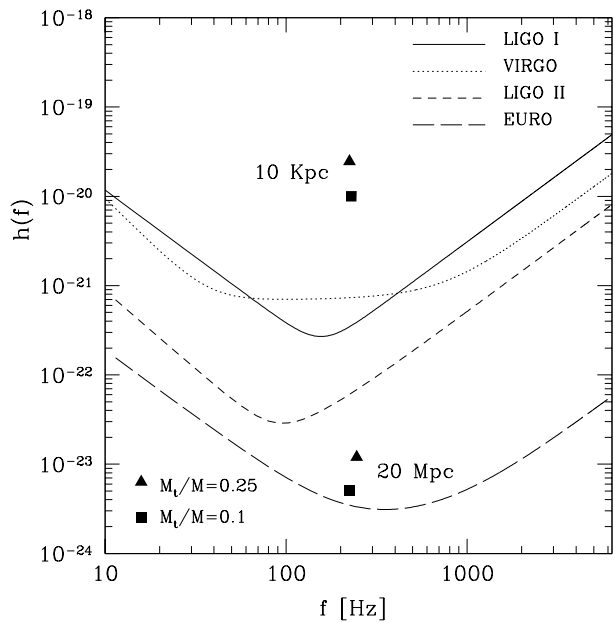,angle=0,width=8.5cm}
        }
\caption{
\label{fig10}
Characteristic wave amplitudes for a perturbed toroidal neutron star with
$\eta=0.04$. These amplitudes have been computed using the strain noise
estimated for LIGO I for a source located at a distance of 10 Kpc and for
LIGO II for a source at a distance of 20 Mpc. The numbers can also be
effectively compared with the sensitivity curves of VIRGO and EURO that
are similar in this frequency range. Different points refer to different
mass ratios, with triangles indicating $M_t/M=0.25$ and squares
$M_t/M=0.1$.}
\end{figure}

\begin{table*}
\begin{center}
\caption{Characteristic properties for the detection of the gravitational
wave signal emitted by a toroidal neutron star. The first two columns
show the mass of toroidal neutron star normalized to the black hole one,
as well as the strength of the initial perturbation $\eta$. The following
nine columns report the characteristic frequency, the characteristic wave
strain and the signal-to-noise ratio computed for the three detectors
LIGO I, LIGO II and EURO, assuming a galactic distance for a source
detected by LIGO I, and an extragalactic distance for a source detected
by LIGO II and EURO. The last column reports the characteristic lifetime
for the existence of toroidal neutron stars that are unstable to the
runaway instability.}
\label{tab2}
\begin{tabular}{ccccccccccccc}
\hline
$M_{\rm t}/M$
       & $\eta$ & $f_c$ (Hz) & $f_c$ (Hz)& $f_c$ (Hz)& $h_c$
       & $h_c$  & $h_c$      & $S/N$     & $S/N$     & $S/N$
       & $\tau_{\rm life}$ (s) \\
       &          &\ LIGO I &\ LIGO II &\ EURO    &\ LIGO I 
       &\ LIGO II &\ EURO   &\ LIGO I  &\ LIGO II &\ EURO &  \\
       &          & (10 Kpc) & (20 Mpc)  & (20 Mpc) & (10Kpc)  & (20 Mpc)
       & (20 Mpc) & (10 Kpc) & (20 Mpc)  & (20 Mpc) &     \\
\hline
$0.50$    & $0.06$ & $234$ &$230$ & $319$ &$5.2\times
10^{-20}$ &$2.6\times 10^{-23}$ &$2.4\times 10^{-23}$ & $98.4$ & $0.4$
& $7.8$ & 0.08 \\
$0.50$ & $0.04$ & $227$ &$223$ &$243$ &$4.3\times
10^{-20}$ &$2.1\times 10^{-23}$ &$2.1\times 10^{-23}$ & $84.4$ & $0.4$
& $6.4$ & 0.12 \\
$0.50$ & $0.02$ & $230$ &$229$ &$238$ &$2.7\times
10^{-20}$ &$1.3\times 10^{-23}$ &$1.3\times 10^{-23}$ & $51.8$  &
$0.2$ & $4.0$ & 0.16 \\
\hline
$0.25$    & $0.06$ & $231$ & $228$ & $248$ & $2.9\times
10^{-20}$ &$1.4\times 10^{-23}$ & $1.4\times 10^{-23}$ & $55.3$ &
$0.2$     & $4.3$  & 0.12 \\
$0.25$ & $0.04$ & $227$ & $224$ & $239$ & $2.4\times
10^{-20}$ &$1.2\times 10^{-23}$ & $1.2\times 10^{-23}$ & $47.4$ &
$0.2$ & $3.6$ & 0.16 \\
$0.25$ & $0.02$ & $229$ & $228$ & $234$ & $1.4\times
10^{-20}$ &$7.2\times 10^{-24}$ & $7.2\times 10^{-24}$& $28.2$  &
$0.1$ & $2.1$ & 0.19 \\
\hline
$0.10$ & $0.06$ & $230$ &$226$ &$250$ & $1.2\times
10^{-20}$ & $6.0\times 10^{-24}$ & $6.0\times 10^{-24}$ &
$23.2$  & $0.1$ &  $1.8$ & 0.14 \\
$0.10$ & $0.04$ & $229$ &$227$ &$239$ & $1.0\times
10^{-20}$ & $5.1\times 10^{-24}$ &$5.0\times 10^{-24}$ & $19.9$
 & $0.1$ & $1.5$ & 0.18 \\
$0.10$ & $0.02$ & $228$ &$227$ &$234$ &$6.1\times
10^{-21}$ &$3.0\times 10^{-24}$ &$3.0\times 10^{-24}$ & $11.9$
 & $<0.1$ & $0.9$ & 0.20 \\
\hline
\end{tabular}
\end{center}
\end{table*}

	Given a detector whose response has a power spectral density
$S_h(f)$, it is then useful to calculate {\it characteristic frequency}
of the signal~\citep{thorne:87}
\begin{equation}
f_c \equiv \left[\int_0^\infty 
	\frac{\langle|{\tilde h}(f)|^2\rangle}{S_h(f)} 
	f df \right]\left[\int_0^\infty
	\frac{\langle|{\tilde h}(f)|^2\rangle}
	{S_h(f)} df \right]^{-1} \ .
\end{equation}
where $\langle|{\tilde h}(f)|^2\rangle$ denotes an average over randomly
distributed angles, that we have simply approximated as $\langle|{\tilde
h}(f)|^2\rangle \simeq |{\tilde h}(f)|^2$. The characteristic frequency
provides a representative measure of where, in frequency, most of the
signal is concentrated and is therefore relevant when the gravitational
wave signal has a rather broad spectrum in frequency, as is the case in a
gravitational core collapse. In the case of an oscillating toroidal
neutron star, on the other hand, the signal is basically
emitted at the fundamental
frequency of oscillation (cf. Table~\ref{tab2}),
increasing the detectability.

	Once the characteristic frequency is known, it can be used to
determine the {\it characteristic amplitude} as
\begin{equation}
\label{h_c}
h_c \equiv \left[3\int_0^\infty \frac{S_h(f_c)}{S_h(f)} 
	\langle|{\tilde h}(f)|^2\rangle f df \right]^{1/2} \ .
\end{equation}
It worth remarking that when the characteristic frequency fits well in
the minima of the sensitivity curves, the weight ${S_h(f_c)}/{S_h(f)}$ in
the integral of (\ref{h_c}) can significantly increase $h_c$ when
compared to $h^{TT}$. This is indeed what happens for the toroidal
neutron stars considered here (cf. Table~\ref{tab2}).

	A direct comparison of characteristic amplitude with the
root-mean-square {\it strain noise} of the detector
\begin{equation}
h_{\rm rms}=\sqrt{f S_h(f)} \ ,
\end{equation}
finally determines the signal-to-noise ratio at the characteristic
frequency as
\begin{equation}
\frac{S}{N} = \frac{h_c}{h_{\rm rms}(f_c)} \ .
\end{equation}

	In Figure~\ref{fig10} we show the characteristic wave amplitude
for sources located at a distance of $R=10$ Kpc and $R=20$ Mpc, as
computed for two different values of the 
toroidal neutron star mass. 
These amplitudes have been computed for the expected
strain noise of LIGO I and LIGO II, but the strain curves of VIRGO
(that is similar in this frequency range,~\citet{damour:01}) and
of EURO~\citep{euro:01} have also been reported for
comparison. Interestingly, with small initial perturbations $(\eta=0.04)$
and mass ratios $(M_{\rm t}/M=0.1)$, the computed characteristic
amplitudes can be above the sensitivity curves of LIGO I for sources
within 10 Kpc and above the sensitivity curve of EURO for sources within
20 Mpc. Both results suggest that a toroidal neutron stars oscillating in
the Virgo cluster could be detectable by the present and planned
interferometric detectors.

	Summarized in Table~\ref{tab2} are the basic properties of the
gravitational wave signal emitted by a toroidal neutron star. Most
notably, we report: the characteristic frequency, the characteristic wave
strain and the signal-to-noise ratios as computed for LIGO I, LIGO II and EURO,
as well as the timescale over which the signal has been computed. All of
these quantities refer to models with different initial perturbations and
located at either 10 Kpc (for LIGO I) or 20 Mpc (for LIGO II and EURO).

	The values for the signal-to-noise ratios reported in
Table~\ref{tab2} are already interesting, but could become larger in at
least three different ways. Firstly, and as remarked above, the signal
strength computed does not include the exponential growth in the
gravitational waveform that would accompany the runaway instability and
that would provide an important contribution to the overall characteristic
amplitude, as it is the case in binary mergers. As an illustrative
example, consider that if the last 10 ms (i.e. roughly the last 5
oscillations) before the disappearence of an unstable torus at 10 Kpc
with $M_{\rm t}/M=0.1$ and an initial perturbation with $\eta=0.02$ were
taken into account, they would yield a final $S/N \simeq 90$. This is to
be contrasted with the corresponding $S/N=11.9$, obtained when the
oscillations are considered on a fixed spacetime (cf. Table~\ref{tab2}).
Secondly, we here have assumed the lifetime of the tori to be limited by
the runaway instability to avoid the uncertainties related to the very
existence of the instability when the specific angular momentum is not
constant. On the other hand, a toroidal neutron star oscillating for a
timescale longer than the one assumed in Table~\ref{tab2} will have a
proportionally stronger signal even when the exponentially growing phase
is neglected. Again, as an example, it is useful to consider that the model
yielding $S/N=11.9$ over 0.2 s would produce a $S/N=16.3$ if $\tau_{\rm
life}=0.4$ s. Thirdly, an oscillating toroidal neutron star stable to the
runaway instability would probably be subject to viscous or magnetic
driven non-axisymmetric instabilities on timescales longer than the ones
discussed here. Once the non-axisymmetric deformations are fully
developed the torus would then have a mass quadrupole with much larger
time variations, losing amounts of energy (and angular momentum) in
gravitational waves larger than the ones computed here.

	The final point to be addressed is the rate at which the emission
of gravitational waves from toroidal neutron stars could be
detected. Although a realistic estimate of this rate is very difficult
since very little is still known about the formation of toroidal neutron
stars, it is reasonable to expect that these objects will be
produced in a significant fraction of the events leading
to core collapse in supernova explosions, 
binary neutron star mergers and tidal disruption of 
neutron stars orbiting a black hole.
Because in
a volume comprising the Virgo cluster all of these sources are among the
most promising ones, overall our results indicate that toroidal neutron
stars could potentially be new sources of gravitational radiation and
certainly suggest a more accurate analysis.

\section{Conclusions}
\label{concl}

	We have performed general relativistic hydrodynamics simulations
of axisymmetric massive and compact tori orbiting a Schwarzschild black
hole with a constant specific angular momentum. These objects have
hydrostatic and hydrodynamical properties (i.e. large masses in small
volumes with central rest-mass densities reaching almost nuclear matter
density) that make them behave effectively as neutron stars, while
possessing a toroidal topology. Our attention has been concentrated on
two different aspects of the dynamics of these tori.

	The first one was focussed on the suggestion that the toroidal
neutron stars might be dynamically unstable to the runaway instability
only if suitably chosen initial data were prescribed. To investigate this
we have used a time-dependent numerical code that integrates the general
relativistic hydrodynamics equations on a curved background using HRSC
methods. The evolution of the spacetime is an essential feature for the
development of the instability and has been modelled through a sequence
of stationary Schwarzschild spacetimes differing only in the total mass
content, which is computed in terms of the total rest-mass accreted onto
the black hole at each instant. The conclusion reached is that, at least
for constant specific angular momentum tori whose self-gravity is
neglected, the initial Roche lobe overflow is not a necessary condition
for the development of the instability, which represents a natural
feature of the dynamics of these objects. Very recent numerical work
shows that this conclusion does not necessarily hold when the specific
angular momentum is not constant~\citep{font2:02}.

	The second aspect was focussed on the dynamical response of these
relativistic tori to the perturbations that are expected to be present
after the catastrophic events that lead to their formation. Upon the
introduction of suitably parametrized perturbations, the toroidal neutron
stars have shown a regular oscillatory behaviour resulting both in a
quasi-periodic variation of the mass accretion rate as well as of the
rest-mass distribution. This response has been interpreted in terms of
the excitation of oscillations modes which could be associated with the
$p$-modes of toroidal neutron stars. These modes, which have been
detected both in their fundamental frequencies as well as in their
overtones, depend systematically on the average density of the tori, and
a disco-seismologic analysis could provide important information on the
physical properties of these toroidal neutron stars.

	High rest-mass densities together with a toroidal topology are
the basic properties that yield large mass quadrupoles for these toroidal
neutron stars. As a consequence of the excitation of oscillations,
the mass quadrupoles are induced to change rapidly and intense
gravitational radiation is thus produced. Estimates made within the
Newtonian quadrupole approximation have shown that strong gravitational
waves can be produced during the short lifetime of these tori.  In
particular, the gravitational radiation emitted by these source is
comparable or larger than the one that is expected during the
gravitational collapse of a stellar iron core, with a rate of detectable events
which could also be larger given the variety of physical scenarios
leading to the formation of a massive torus orbiting a black
hole. Overall, the strength of the gravitational waves emitted and their
periodicity are such that signal-to-noise ratios $\sim {\cal O}(1)-{\cal
O}(10)$ can be reached for sources at 20000 or 10 Kpc respectively,
making these new sources of gravitational waves detectable and
potentially important.

	The results reported here are  a first step towards the
understanding of the dynamics of toroidal neutron stars and call for a
number of natural extensions and improvements.
In particular, we plan to extend the simulations to the
more realistic scenario of non-constant angular momentum tori, in order
to find out if our conclusions regarding both the oscillation properties
of toroidal neutron stars 
and the large amplitude of the associated gravitational wave
emission still hold. In addition, it is interesting to improve, using
more accurate approaches, the calculation of the gravitational radiation
from the oscillations as well as from the runaway instability in the
region of the parameter space where the instability could
exist~\citep{font2:02}. Equally interesting is to include the effects
introduced by self-gravity of the torus and determine the quantitative
differences that will be encountered in this case. Last but not least, it
is worth analyzing how the presence of a Kerr black hole would modify the
present results, determining, in particular, the dependence of this
phenomenology on the spin of the central black hole.

	The results obtained here also provide promising suggestions for
further work in a number of research areas different from the ones
considered here. Firstly, they  indicate that more needs to
be investigated about the oscillation properties of
relativistic tori and geometrically thick discs. 
Secondly, they suggest the possibility that the
phenomenology observed in the quasi-periodic X-ray luminosity existing in
LMXB's might be related, in some form, to the quasi-periodic accretion
resulting from oscillation modes of the discs~\citep{markovic:98,
psaltis:00, wagoner:01}. Thirdly, and perhaps most importantly, these
results show that new and unexpected sources of gravitational radiation
could exist and might be observed when the new detectors of gravitational
radiation become fully operative in the near future.

\section*{Acknowledgments}
It is a pleasure to thank M. Abramowicz, F. Daigne, J.M. Ib\'a\~nez,
J. Miller, E. M\"uller and S'i.  Yoshida for many interesting and
insightful discussions. Special thanks go to N. Andersson and
B. Sathyaprakash for their comments on the detectability of the
gravitational wave signal. L.R. also acknowledges useful discussions with
V. Ferrari, J. Friedman, B. Schutz, L. Stella and K. Uryu.  Financial
support for this research has been provided by the MIUR and by the EU
Network Programme (Research Training Network Contract
HPRN-CT-2000-00137).  J.A.F.  acknowledges financial support from a EU
Marie Curie fellowship (HPMF-CT-2001-01172) and from the Spanish
Ministerio de Ciencia y Tecnolog\'{\i}a (grant AYA 2001-3490-C02-01). The
computations were performed on the Beowulf Cluster for numerical
relativity {\it ``Albert100''}, at the University of Parma.

\bibliographystyle{mn2e}
\bibliography{paper}

\begin{thebibliography}{}

\bibitem[\protect\citeauthoryear{{Abramowicz}}{{Abramowicz}}{1974}]
{abramowicz:74} {Abramowicz} M.~A., 1974, Acta Astron., 24, 45

\bibitem[\protect\citeauthoryear{{Abramowicz},{Jaroszy\'nski},{Sikora}}
{{Abramowicz et~al.}}{1978}]{abramowicz:78} {Abramowicz}
M.~A. {Jaroszy\'nski} M. \& {Sikora} M., 1978, A\&A, 63, 221.

\bibitem[\protect\citeauthoryear{{Abramowicz}, {Calvani} \&
{Nobili}}{{Abramowicz} et~al.}{1983}]{abramowicz:83} {Abramowicz} M.~A.,
{Calvani} M., {Nobili} L., 1983, Nature, 302, 597

\bibitem[\protect\citeauthoryear{{Abramowicz}, {Karas} \&
{Lanza}}{{Abramowicz} et~al.}{1998}]{abramowicz:98} {Abramowicz}
M.~A., {Karas} V., {Lanza} A., 1998, A\&A, 331, 1143

\bibitem[\protect\citeauthoryear{{Ansorg}, {Kleinw\"acther} \&
{Meinel}}{{Ansorg} et~al.}{2002}]{ansorg:02} {Ansorg}
M., {Kleinw\"acther} A., {Meinel} R., 2002, {\tt astro-ph/0208267}


\bibitem[\protect\citeauthoryear{{Banyuls}, {Font}, {Ib\'a\~nez},
{Mart\'{\i}} \& {Miralles}}{{Banyuls} et~al.}{1997}]{banyuls:97}
{Banyuls} F., {Font} J.~A., {Ib\'a\~nez} J.~M., {Mart\'{\i}} J.~M.,
{Miralles} J.~A., 1997, ApJ, 476, 221


\bibitem[\protect\citeauthoryear{{Blanchet}, {Damour} \&
{Sch\"afer}}{{Blanchet} et~al.}{2001}]{blanchet:90}{Blanchet} L.,
{Damour} T., {Sch\"afer} G. 1990, MNRAS, 242, 289


\bibitem[\protect\citeauthoryear{{Boyer}}{{Boyer}}{1965}]{boyer:65}
{Boyer} R.~H., 1965, Proc. Cambridge Phil. Soc. 61, 527

\bibitem[\protect\citeauthoryear{{Brandt}, Font, Ib\'a\~nez, Mass\'o \&
{Seidel}}{{Brandt} et~al.}{2000}]{brandt:00} {Brandt} S., Font J.~A.,
Ib\'a\~nez J.~M., Mass\'o J., {Seidel} E., 2000, Comput. Phys. Comm.,
124, 169


\bibitem[\protect\citeauthoryear{{Daigne} \& {Mochkovitch}}{{Daigne} \&
{Mochkovitch}}{1997}]{daigne:97} {Daigne} F., {Mochkovitch} R., 1997,
MNRAS, 285, L15

\bibitem[\protect\citeauthoryear{{Damour}, {Iyer} \&
{Sathyaprakash}}{{Damour} et~al.}{2001}]{damour:01}{Damour} T., {Iyer}
B.~R., {Sathyaprakash} B.~S. 2001, Phys. Rev. D, 63, 044023

\bibitem[\protect\citeauthoryear{{Dimmelmeier}, Font, M\"uller}
{{Dimmelmeier} et~al.}{2002}]{dimmel:02} {Dimmelmeier} H., {Font} J.~A.,
{M\"uller} E., 2002, A\&A, 393, 523

\bibitem[\protect\citeauthoryear{{Donat} \& {Marquina}}{{Donat} \&
{Marquina}}{1996}]{donat:96} {Donat} R., {Marquina} A., 1996,
J. Comp. Phys., 125, 42




\bibitem[\protect\citeauthoryear{{Finn}}{1989}]{finn:89}
{Finn} L. S. in {\it Frontiers in Numerical Relativity},
ed. C.R.~Evans, S. L.~Finn., \& D.W.~Hobill,
Cambridge University Press, Cambridge, England, 1989

\bibitem[\protect\citeauthoryear{{Fishbone} \& {Moncrief}}{{Fishbone} \&
{Moncrief}}{1976}]{fishbone:76} {Fishbone} L.~G., {Moncrief} V., 1976,
ApJ, 207, 962

\bibitem[\protect\citeauthoryear{{Font}}{{Font}}{2000}]{fontlr}
{Font} J.~A.,  2000, Living Reviews in Relativity, 3, 2

\bibitem[\protect\citeauthoryear{{Font}, Goodale, Iyer, Miller, Rezzolla,
Seidel, Stergioulas, Suen \& Tobias}{{Font} et~al.}{2002}]{font-goodale:02}
{Font} J.~A., Goodale T., Iyer S., Miller M., Rezzolla L., Seidel E.,
Stergioulas N., Suen W., Tobias M., 2002, Phys. Rev. D 65, 084024

\bibitem[\protect\citeauthoryear{{Font} \& {Ib\'a\~nez}}{{Font} \&
{Ib\'a\~nez}}{1998a}]{font:98a} {Font} J.~A., {Ib\'a\~nez} J.~M., 1998a,
ApJ, 494, 297




\bibitem[\protect\citeauthoryear{{Font} \& {Daigne}}{{Font} \&
{Daigne}}{2002a}] {font:02}{Font} J.~A., {Daigne} F., 2002a, MNRAS, 334,
383

\bibitem[\protect\citeauthoryear{{Font} \& {Daigne}}{{Font} \&
{Daigne}}{2002b}] {font2:02}{Font} J.~A., {Daigne} F., 2002b, ApJL, {\it 
submitted} 


\bibitem[\protect\citeauthoryear{{Hawley}, {Wilson} \& {Smarr}}{{Hawley}
et~al.}{1984a}]{hawley:84a} {Hawley} J.~F., {Wilson} J.~R., {Smarr}
L.~L., 1984a, ApJ, 277, 296


\bibitem[\protect\citeauthoryear{{Igumenshchev} \&
{Beloborodov}}{{Igumenshchev} \& {Beloborodov}}{1997}]{igumenshchev:97}
{Igumenshchev} I.~V., {Beloborodov} A.~M., 1997, MNRAS, 284, 767





\bibitem[\protect\citeauthoryear{{Kato}}{{Kato}}{2001}]{kato:01}
{Kato} S., 2001, PASJ, 53, 1

\bibitem[\protect\citeauthoryear{{Kochanek}, {Shapiro} \&
{Teukolsky}{Chernoff}}{{Kochanek} et~al.}{1990}]{kochanek:90} {Kochanek} C.,
{Shapiro} S.~L., {Teukolsky} S.~A. {Chernoff} D.~F.,
1990, ApJ, 358, 81

\bibitem[\protect\citeauthoryear{{Kozlowski}, {Jaroszynski} \&
{Abramowicz}}{{Kozlowski} et~al.}{1978}]{kozlowski:78} {Kozlowski} M.,
{Jaroszynski} M., {Abramowicz} M.~A., 1978, A\&A, 63, 209

\bibitem[\protect\citeauthoryear{{Lee} \& {Kluzniak}}{{Lee} \&
{Kluzniak}}{1999a}] {lee:99a}{Lee} W.~H., {Kluzniak} W., 1999a,
MNRAS, 308, 780

\bibitem[\protect\citeauthoryear{{Lee} \& {Kluzniak}}{{Lee} \&
{Kluzniak}}{1999b}] {lee:99b}{Lee} W.~H., {Kluzniak} W., 1999b,
ApJ, 526, 178

\bibitem[\protect\citeauthoryear{{Lee}}{{Lee}}{2000}] {lee:00}{Lee}
W.~H., MNRAS, 318, 606




\bibitem[\protect\citeauthoryear{{MacFadyen} \& {Woosley}}{{MacFayden} \&
{Woosley}}{1999}] {macfayden:99}{MacFayden} A.~I., {Woosley} S.~E., 1999,
ApJ, 524, 262

\bibitem[\protect\citeauthoryear{{Markovic} \& {Lamb}}{{Markovic} \&
{Lamb}}{1998}] {markovic:98}{Markovic} D., {Lamb} F.~K., 1998, ApJ, 
507, 316

\bibitem[\protect\citeauthoryear{{Masuda} \& {Eriguchi}}{{Masuda} \&
{Eriguchi}}{1997}]{masuda:97} {Masuda} N., {Eriguchi} Y., 1997, ApJ, 489,
804

\bibitem[\protect\citeauthoryear{{Masuda}, {Nishida} \&
{Eriguchi}}{{Masuda} et~al.}{1998}]{masuda:98} {Masuda}
N., {Nishida} S., {Eriguchi} Y., 1998, MNRAS, 297, 1139

\bibitem[\protect\citeauthoryear{Meszaros}{Meszaros}{2002}]{meszaros:02}
{Meszaros} P., 2002, ARA\&A, 40, 137



\bibitem[\protect\citeauthoryear{Michel}{Michel}{1972}]{michel:72} Michel
F., 1972, Astrophys. Spa. Sci., 15, 153

\bibitem[\protect\citeauthoryear{Mineshige, Hosokawa, Machida \&
Matsumoto}{Mineshige et~al.}{2002}]{mineshige:02}{Mineshige} S., {Hosokawa}
T., {Machida} M., {Matsumoto} R. 2002, {\tt astroph/0208406}




\bibitem[\protect\citeauthoryear{{Novikov} \& {Thorne}}{{Novikov} \&
{Thorne}}{1973}]{novikov:73} {Novikov} I., {Thorne} K.~S., 1973, in {\it
Black Holes}, eds. B. de Witt and C. de Witt, Gordon and Breach, NY
491, 663


\bibitem[\protect\citeauthoryear{{Perez}, {Silbergleit}, {Wagoner} \&
{Lehr}}{{Perez} et~al.}{1997}]{perez:97} {Perez} C.~A., {Silbergleit}
A.~S., {Wagoner} R.~V., Lehr D.~E., 1997, ApJ, 476, 589 

\bibitem[\protect\citeauthoryear{{Psaltis}, \& {Norman}}{{Psaltis} \&
{Norman}}{2000}]{psaltis:00} {Psaltis} D., {Norman} C., {\tt
astro-ph/0001391}







\bibitem[\protect\citeauthoryear{{Rezzolla}, {Shibata},
{Asada}, {Baumgarte} \& {Shapiro}}{{Rezzolla}
et~al.}{1999}]{rezzolla:99} {Rezzolla} L., {Shibata} M., 
{Asada} H., {Baumgarte} T.~W. {Shapiro} S.~L., 1999 ApJ,
525, 935

\bibitem[\protect\citeauthoryear{{Rezzolla}, {Lamb} \&
{Shapiro}}{{Rezzolla} et~al.}{2000}]{rls:00} {Rezzolla} L., {Lamb}
F.~K., {Shapiro} S.~L., 2000 ApJL, 531, L141


\bibitem[\protect\citeauthoryear{{Rezzolla}, {Yoshida} \&
{Zanotti}}{{Rezzolla} et~al.}{2002}]{ryz:02} {Rezzolla} L., {Yoshida}
S'i., {Zanotti} O., 2002 {\it in preparation}

\bibitem[\protect\citeauthoryear{{Rodriguez}, {Silbergleit} \&
{Wagoner}}{{Rodriguez} et~al.}{2002}]{rodriguez:02} {Rodriguez} M.~O.,
{Silbergleit} A.~S., {Wagoner} R.~V., 2002, ApJ, 567, 1043 

\bibitem[\protect\citeauthoryear{{Ruffert} \& {Janka}}{{Ruffert} \&
{Janka}}{1999}]{ruffert:99} {Ruffert} M., {Janka} H.-T., 1999, A\&A, 344,
573

\bibitem[\protect\citeauthoryear{{Ruffert} \& {Janka}}{{Ruffert} \&
{Janka}}{2001}]{ruffert:01} {Ruffert} M., {Janka} H.-T., 2001, A\&A, 380,
544


\bibitem[\protect\citeauthoryear{{Shakura} \& {Sunyaev}}{{Shakura} \&
{Sunyaev}}{1973}]{shakura:73} {Shakura} N. I., {Sunyaev}
R. A., 1973, A\&A, 24, 337


\bibitem[\protect\citeauthoryear{{Shibata} \& {Ury{\= u}}}{{Shibata} \&
{Ury{\= u}}}{2000}]{shibata:00} {Shibata} M., {Ury{\= u}} K., 2000,
Phys. Rev. D, 61, 4001

\bibitem[\protect\citeauthoryear{{Shibata}}{{Shibata}}{2002}]{shibata:02}
{Shibata} M., 2002, Talk given at the IV Meeting of the European Network
on {\it ``Sources of Gravitational waves''}, Palma de Mallorca, Spain

\bibitem[\protect\citeauthoryear{{Silbergleit}, {Wagoner} \&
{Rodriguez}}{{Silbergleit} et~al.}{2001}]{silbergleit:01} {Silbergleit}
A.~S., {Wagoner} R.~V., {Rodriguez} M.~O., 2001, ApJ, 548, 335 

\bibitem[\protect\citeauthoryear{{Spruit}}{{Spruit}}{1999}]{spruit:99}
{Spruit} H.~C., 1999, A\&A, 341,  L1 

\bibitem[\protect\citeauthoryear{{Thorne}}{1987}]{thorne:87}
{Thorne} K. S. in {\it 300 Years of Gravitation},
edited by S. W. Hawking ans W. Israel, Cambridge
University Press, Cambridge, England, 1987

\bibitem[\protect\citeauthoryear{{Wagoner}, {Silbergleit}, \&
{Rodriguez}}{{Wagoner} et~al.}{2002}]{wagoner:01} {Wagoner} R.~V.,
{Silbergleit} A.~S., {Rodriguez} M.~O., 2001, ApJ, 559, L25

\bibitem[\protect\citeauthoryear{{Wilson}}{{Wilson}}{1984}]{wilson:84}
{Wilson} D.~B., 1984, Nature, 312, 620




\bibitem[\protect\citeauthoryear{{van der Klis}}{{van der Klis}}{2000}]
{van_der_klis:00} {van der Klis} M., 2000, ARA\&A, 38, 71

\bibitem[\protect\citeauthoryear{{van Putten}}{{van Putten}}{2001a}]
{van_putten:01a} {van Putten} M.~H.~P.~M., 2001, Phys. Rev. Lett. 87,
091101

\bibitem[\protect\citeauthoryear{{van Putten}}{{van Putten}}{2001b}]
{van_putten:01b} {van Putten} M.~H.~P.~M., 2001, ApJ. 575, L71

\bibitem[\protect\citeauthoryear{Vietri \& Stella}{Vietri \& Stella} {1998}]
{vietri_e_stella:98} {Vietri} M., {Stella} L. 1998, ApJ. 507, L45

\bibitem[\protect\citeauthoryear{Winkler}{Winkler} {2001}] {euro:01}
{Winkler W., for ``The VIRGO/GEO Collaboration'', Vision document} {\tt
www.astro.cf.ac.uk/geo/euro/}



\bibitem[\protect\citeauthoryear{{Zwerger} \& {M\"uller}}{{Zwerger} \&
{M\"uller}}{1997}]{zwerger:97} {Zwerger} T., {M\"uller}
E., 1997, A\&A, 320, 209

\end{thebibliography}

\label{lastpage}  
\end{document}